\documentclass[journal ]{new-aiaa}
\usepackage[utf8]{inputenc}
\usepackage{textcomp}
\usepackage{graphicx}
\usepackage{amsmath}
\usepackage[version=4]{mhchem}
\usepackage{siunitx}
\usepackage{longtable,tabularx}
\usepackage[flushleft]{threeparttable}
\usepackage{booktabs}
\usepackage{gensymb}
\usepackage{mdframed}
\usepackage{subcaption}
\usepackage{tcolorbox}
\usepackage{bm}
\let\svthefootnote\thefootnote
\newcommand\freefootnote[1]{
  \let\thefootnote\relax
  \footnotetext{#1}
  \let\thefootnote\svthefootnote
}

\definecolor{myblue}{rgb}{0, 0.23, 0.64}
\definecolor{WVUblue}{rgb}{0, 0.16, 0.33}
 
\hypersetup{
    colorlinks=true,
    linkcolor=myblue,
    filecolor=magenta,
    citecolor=myblue,
    urlcolor=myblue,
}

\title{Benchmarking Agility and Reconfigurability in Satellite Systems for Tropical Cyclone Monitoring}

\author{Brycen D. Pearl\footnote{Ph.D. Student, Department of Mechanical, Materials and Aerospace Engineering, Student Member AIAA.}, Logan P. Gold\footnote{Undergraduate Student, Department of Mechanical, Materials and Aerospace Engineering.}, and Hang Woon Lee\footnote{Assistant Professor, Department of Mechanical, Materials and Aerospace Engineering; hangwoon.lee@mail.wvu.edu. Member AIAA (Corresponding Author).}}
\affil{West Virginia University, Morgantown, WV, 26506}

\begin{document}

\newpage

\freefootnote{This paper is a substantially revised version of the Paper AAS 23-191, presented at the AAS/AIAA Astrodynamics Specialist Conference, Big Sky, MT, August 13-17, 2023. It offers new results and a better description of the materials.}

\maketitle

\begin{abstract} 
Tropical cyclones (TCs) are highly dynamic natural disasters that travel vast distances and occupy a large spatial scale, leading to loss of life, economic strife, and destruction of infrastructure. The severe impact of TCs makes them crucial to monitor such that the collected data contributes to forecasting their trajectory and severity, as well as the provision of information to relief agencies. Among the various methods used to monitor TCs, Earth observation satellites are the most flexible, allowing for frequent observations with a wide variety of instruments. Traditionally, satellite scheduling algorithms assume nadir-directional observations, a limitation that can be alleviated by incorporating satellite agility and constellation reconfigurability---two state-of-the-art concepts of operations (CONOPS) that extend the amount of time TCs can be observed from orbit. This paper conducts a systematic comparative analysis between both CONOPS to present the performance of each relative to baseline nadir-directional observations in monitoring TCs. A dataset of 100 historical TCs is used to provide a benchmark concerning real-world data through maximizing the number of quality observations. The results of the comparative analysis indicate that constellation reconfigurability allowing plane-change maneuvers outperforms satellite agility in the majority of TCs analyzed.
\end{abstract}

\section*{Nomenclature}
{\renewcommand\arraystretch{1.0}
\noindent\begin{longtable*}{@{}l @{\quad=\quad} l@{}}
\multicolumn{2}{@{}l}{Parameters}\\
$c$ & Cost of orbital maneuver \\
$\bm{D}$ & Satellite pointing direction\\
$J$ & Number of orbital slots \\ 
$K$ & Number of satellites \\
$M$ & Rotation matrix \\
$\bm{N}$ & Satellite nadir pointing direction \\
$P$ & Number of targets for observation \\
$\bm{P}$ & Target position vector \\
$\bm{r}$ & Satellite position vector \\ 
$S$ & Number of stages for reconfiguration \\
$T$ & Number of time steps \\
$\bm{T}$ & Target pointing direction \\
$T_{\text{r}}$ & Mission duration \\
$T_{\text{s}}$ & Number of time steps in each stage of reconfiguration \\ 
$V$ & Visibility of a target from satellite orbital slot \\
$x$ & Decision variable for the reconfiguration path of a satellite \\
$y$ & Indicator variable for visibility state \\
$z$ & Sum of observation rewards \\
$\alpha$ & Decision variable for Euler angle rotation about $\hat{x}$ \\
$\beta$  & Decision variable for Euler angle rotation about $\hat{y}$ \\
$\gamma$ & Decision variable for Euler angle rotation about $\hat{z}$ \\
$\Delta t$ & Time step size between discrete time steps \\
$\Delta \tau$ & Time step size between attitude control opportunities \\
$\zeta$ & Euler angle rotation limit \\
$\theta$ & Angular difference in pointing directions \\
$\pi$ & Obtainable observation reward matrix \\
\multicolumn{2}{@{}l}{Sets}\\
$\mathcal{J}$ & Set of orbital slots \\
$\mathcal{K}$ & Set of satellites \\
$\mathcal{P}$ & Set of targets \\
$\mathcal{S}$ & Set of stages for reconfiguration \\
$\mathcal{T}$ & Set of time steps \\
$\mathcal{T}^\prime$ & Set of attitude control opportunities \\
$\mathcal{T}_{\text{s}}$ & Set of stage horizon time steps \\
\multicolumn{2}{@{}l}{Subscripts and indexing}\\
$i, j$ & Orbital slot indices \\
$k$ & Satellite index \\
$p$ & Target index \\
$s$ & Stage index \\
$t$ & Time step index \\
${\tau}$ & Control opportunity index \\
\end{longtable*}}

\section{Introduction} \label{sec:introduction}

\lettrine{T}{ropical} cyclones (TCs) develop as a result of atmospheric convection caused by a differential in temperature, being fueled by warm sea surface temperatures and oceanic currents \cite{Gierach2007}. TCs are also highly dynamic and span a significant spatial scale from \num{50} to \SI{1300}{km}, most frequently falling between a radial distance of \num{700} and \SI{800}{km} \cite{Perez2021size}, in addition to traveling vast distances. The large range that TCs cover leads to widespread economic strife, damaged infrastructure, and loss of life, such as in the cases of Hurricanes Katrina and Sandy, among many others. Hurricane Katrina, for example, made landfall along the US Golf Coast in 2005 as a Category 3 hurricane, causing roughly \num{986} total deaths either directly or indirectly \cite{Brunkard2008}, flooding an estimated \SI{80}{\%} of New Orleans \cite{Brunkard2008}, and causing approximately \$156 billion in damage \cite{Burton2005}. Similarly, Hurricane Sandy made landfall along the US northeastern coastline in 2012 as a Category 3 hurricane \cite{Lockyer2018}, causing roughly \num{233} total deaths \cite{Diakakis2015}, and causing between \$78 and \$97 billion in damage \cite{Kunz2013}. As a result of the impact of TCs, forecasting the trajectory and severity of TCs proves crucial, providing information to relief agencies and ideally an adequate lead time for response. 

A wide variety of methods are used to gather information on TCs, including land-based facilities, aircraft, and airborne platforms, all of which can be equipped with many instruments such as radar, lidar, radiometers, and pressure, temperature, and humidity (PTH) sensors. Land-based sensors employ Doppler radar instruments to measure energy and calculate TC wind fields, which are useful in assessing the maximum wind speed and directionality of TCs, allowing a prediction of not only the location of landfall but also the potential damage upon landfall \cite{Raghavan2013}. Similarly, aircraft are utilized to fly adjacent to TCs while carrying Stepped-Frequency Microwave Radiometers and disposable airborne probes that allow the in-situ measurement of pressure, temperature, humidity, as well as wind speed \cite{HOLBACH2023}. Additionally, the Hurricane Imaging Radiometer collects in-flight measurements on the wind speed and rain rate of TCs through the use of synthetic aperture thinned array radiometry \cite{Amarin2012}. Airborne sensors, such as dropsondes released by aircraft, are expendable probes equipped with PTH sensors and GPS systems that can be used to track TCs when deployed within the storm \cite{HOLBACH2023}. 

In conjunction, Earth observation (EO) satellites in low Earth orbit (LEO), equipped with a variety of unique sensors, are used to gather data regarding sea surface temperatures, subsurface water response, sea surface height, and chlorophyll levels, among others. For example, the Compact Ocean Wind Vector Radiometer (COWVR), designed to gather ocean vector winds from small satellite platforms \cite{COWVR-7943884}, allows tracking and computation related to ocean vector winds that influence the directionality, and often intensity, of TCs. Additionally, the Temporal Experiment for Storms and Tropical Systems Technology (TEMPEST), designed to gather precipitation and ice accumulation within clouds \cite{TEMPEST-8517330}, allows the tracking of intense precipitation within TCs. Furthermore, the Moderate Resolution Imaging Spectroradiometer (MODIS), designed to capture optical imagery of various Earth-based planetary phenomena \cite{Wilson2018MODIS}, is used to image TCs, contributing to the chlorophyll levels utilized in the study by Ref.~\cite{Gierach2007}. Finally, dedicated satellite missions such as the Cyclone Global Navigation Satellite System, a constellation of eight satellites, also measure ocean surface wind fields at a rate higher than that of previous systems \cite{Rose2013}. As a result of the flexibility provided by satellite systems, they are regarded as imperative in monitoring TCs, among many other natural disasters \cite{boustan2020}. 

The primary limiting factor in satellite observations is present as a result of the \textit{visible time window} (VTW), defined as the time during which a target is visible to a satellite \cite{SUN2019EOSSET,CHATTERJEE2022}. As such, maximizing the VTW of a TC directly correlates to the maximization of data collected, which is highly important due to the devastating effects of TCs that may be mitigated if more data is provided. The VTW is a key parameter in satellite scheduling algorithms, as observations of a target may only be scheduled based upon the VTW of the target. Traditional EO satellite scheduling problems assume standard nadir-directional observations \cite{Rose2013}, thus considering a constant VTW \cite{Chen2020,CHATTERJEE2022}. State-of-the-art satellite concepts of operations (CONOPS) enable satellites to manipulate the VTW of targets through various means. For example, \textit{satellite agility}, defined as the ability of satellites to perform attitude control maneuvers (slewing) in roll, pitch, and yaw \cite{LEMAITRE2002}, can extend the VTW or create new VTW opportunities through slewing beyond the nadir swath width \cite{CHATTERJEE2022}. Additionally, \textit{satellite maneuverability}, defined as the ability of satellites to perform orbital maneuvers, and \textit{constellation reconfigurability}, defined as the ability of a constellation of maneuverable satellites to reconfigure from a given configuration to a more optimal configuration, can also extend the VTW or create new VTW opportunities through the more optimal reconfigured constellation.

\begin{figure}[!ht]
    \centering
    \begin{subfigure}[h]{0.33\textwidth}
        \centering
        \includegraphics[width = \textwidth]{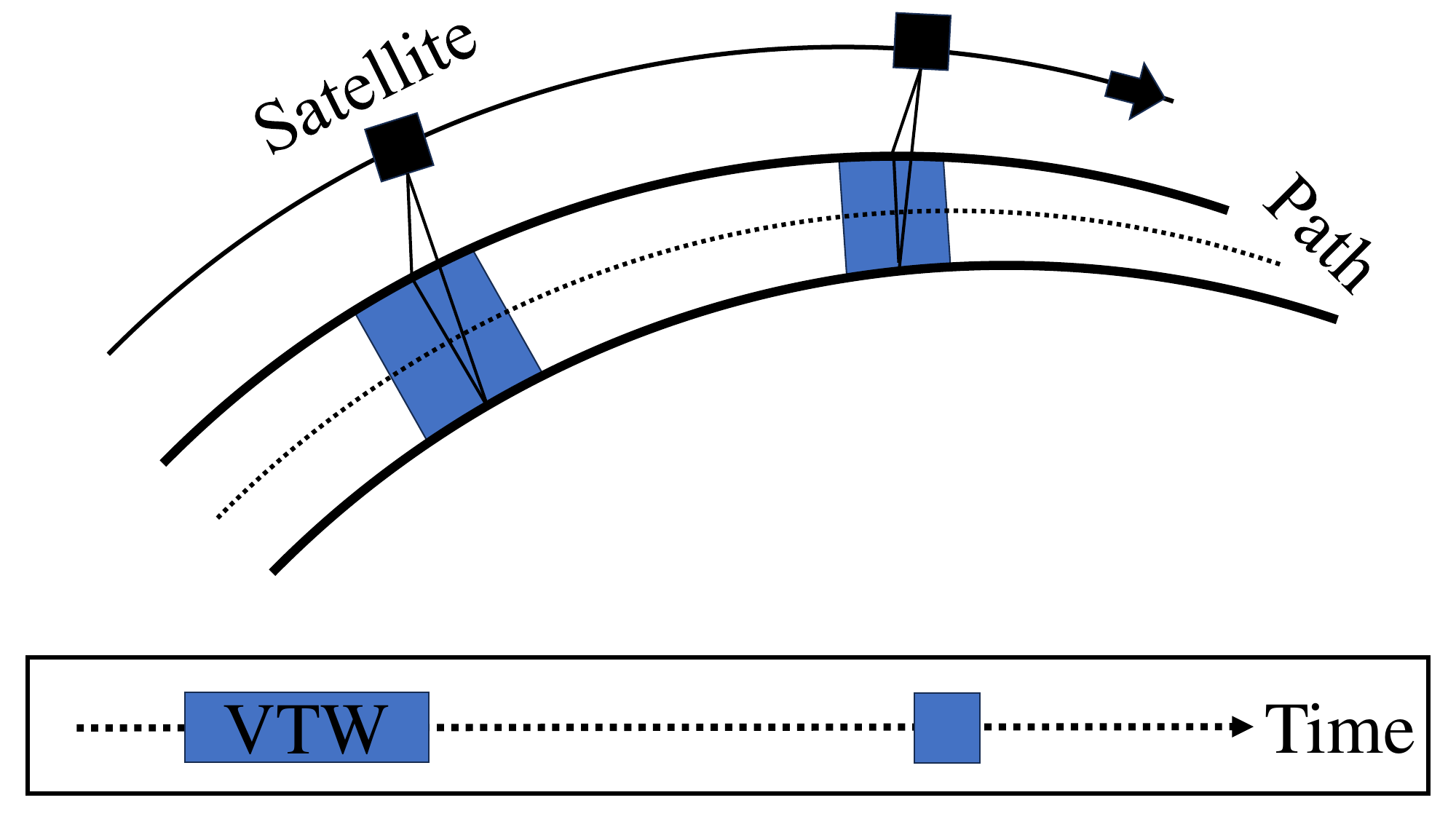}
        \caption{Nadir directional (baseline).}
        \label{fig:nadir_FOV}
    \end{subfigure}
    \begin{subfigure}[h]{0.33\textwidth}
        \centering
        \includegraphics[width = \textwidth]{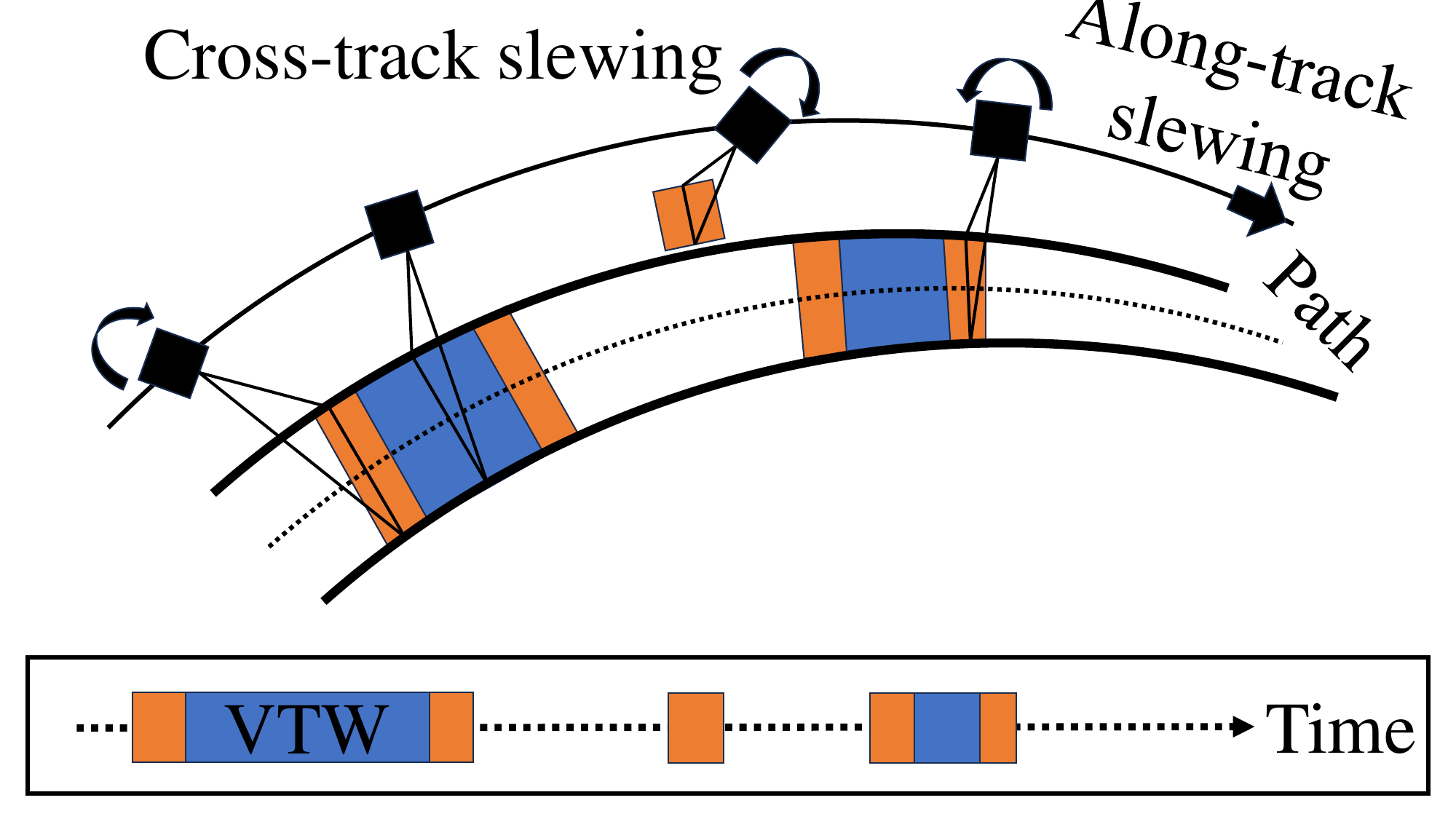}
        \caption{Satellite agility.}
        \label{fig:agile_FOV}
    \end{subfigure}
    \begin{subfigure}[h]{0.33\textwidth}
        \centering
        \includegraphics[width = \textwidth]{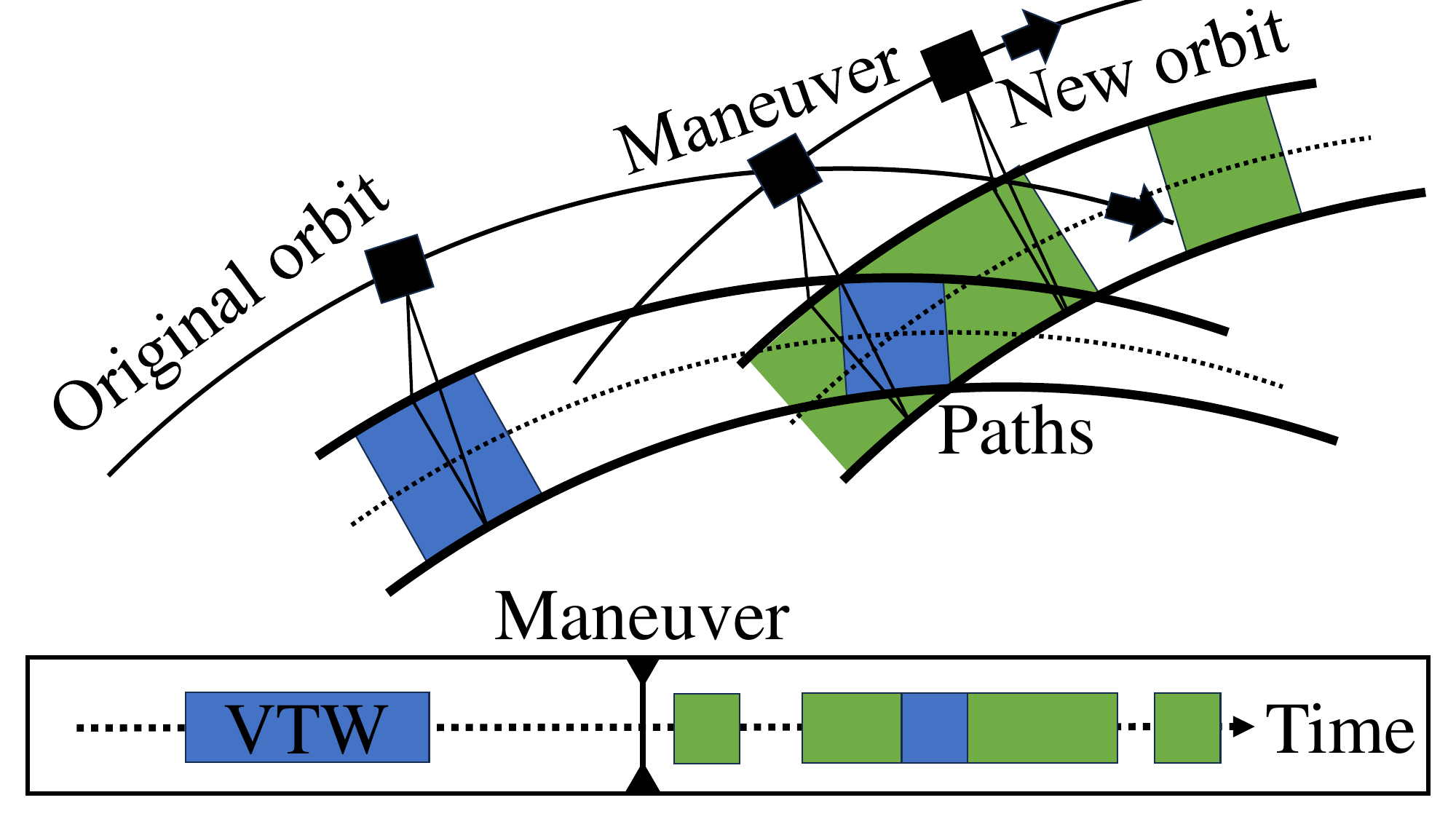}
        \caption{Satellite maneuverability.}
        \label{fig:recon_FOV}
    \end{subfigure}
    \caption{Difference in VTW of CONOPS.}
    \label{fig:VTW}
\end{figure}

While satellite agility and constellation reconfigurability both have the ability to manipulate the VTW, they do so in a different manner, as highlighted and demonstrated in Fig.~\ref{fig:VTW}. The figure depicts the VTW resulting from an arbitrary target existing within a field of view (FOV) and provides a nadir-directional VTW as a baseline. The concept of satellite agility shows an extension to the VTW in orange, resulting from slewing in the along-track and cross-track directions. Similarly, the concept of satellite maneuverability shows a new satellite orbit post-maneuver with the associated new VTWs in green. 

Despite the potential to extend the VTW to improve system flexibility, each CONOPS has limitations as a result of assumptions made by proposed formulations. Satellite agility presents a degradation in optical observations due to optical tilt and change of resolution \cite{Song2018} resulting from a change in the look angle caused by slewing, where observations at the nadir are of higher quality than angled observations \cite{Peng2020timedependent}. While Refs.~\cite{Song2018,Peng2020timedependent,CHATTERJEE2022} apply this degradation for general targets, it follows logically that optical observations of TCs would also apply such a degradation. Additionally, assumptions present in previous constellation reconfigurability research, such as simple analytical cost computation algorithms \cite{lee2024deterministic} or the restriction to only altitude maneuvers \cite{Morgan2023}, limit the full potential due to the high cost of orbital maneuvers in LEO or a lowered flexibility, respectively. Certain key implementations can improve upon these limitations, such as the utilization of trajectory optimization to decrease maneuver costs, and incorporating a large number of transfer options for each satellite within a constellation to improve overall flexibility. Additionally, expanding upon previously conducted multi-stage constellation reconfigurability research \cite{Lee2022,Lee2023,lee2024deterministic} can improve upon single-stage constellation reconfigurability demonstrated in \cite{Lee2020binary,Lee2021lagrangian,lee2023regional}.

This paper presents a systematic comparative analysis of satellite agility and constellation reconfigurability in EO, evaluating their respective merits and benchmarking them against a baseline nadir-directional case; this comparative analysis, to the best of the author's knowledge, is the first of its kind. Through this analysis, several key objectives are accomplished. Firstly, both satellite agility and constellation reconfigurability have been compared in respective research to a baseline nadir-directional system \cite{Branch2023,Morgan2023,lee2024deterministic}, but they have yet to be compared to one another to evaluate their relative merits. Additionally, while agility has seen extensive research, constellation reconfigurability is still an emergent CONOPS worthy of further investigation in an environment comprised of a large set of real-world target data. Secondly, while satellite agility has seen implementation in real-world systems, such as the NASA Distributed Spacecraft with Heuristic Intelligence to Enable Logistical Decisions (D-SHIELD) onboard agile scheduling algorithm \cite{Nag2020}, the CNES Pl\'eiades agile two-satellite constellation \cite{LEMAITRE2002}, and the WorldView agile four-satellite constellation \cite{Wang2021}, constellation reconfigurability has yet to be implemented for EO. A rigorous comparison of satellite agility and constellation reconfigurability, building upon the authors' previous research \cite{Pearl2023}, will present the potential value of reconfigurability in EO, especially with respect to highly dynamic TCs. The results of the comparative analysis, as will be shown later in this paper, indicate that constellation reconfigurability and satellite agility both have interesting cases in which one or the other has a higher level of performance. Additionally, the results indicate that constellation reconfigurability is a promising CONOPS that may outperform satellite agility when given adequate operational parameters.

The rest of the paper is organized as follows. Section~\ref{sec:constellation_models} describes the details of the modeling methodology for each CONOPS. Then, Sec.~\ref{sec:comparative_analysis} discusses any relevant parameters given to each CONOPS and the TC data used for analysis, followed by presenting the results of simulating each CONOPS model with respect to the TC data. Finally, Sec.~\ref{sec:conclusion} provides commentary on the results and conclusions, then provides direction for future research.

\section{Concepts of Operations} \label{sec:constellation_models}

Mathematical models are created to compare the performance of each CONOPS. Standard nadir-directional operation is modeled as a constellation of nadir-directional satellites and is referred to as the \textit{baseline model}. Separately, satellite agility and constellation reconfigurability are formulated to maximize the number of quality observations, represented as the sum of observation rewards, $z$, with decision variables related to the associated CONOPS. It should be noted that this comparison considers optical observations rather than other forms of measurements, and as such, observation reward degradation will be incorporated for satellite agility. The concept of satellite agility incorporates slewing, allowing the modeled satellites to point in the direction of the TCs. Separately, the concept of constellation reconfigurability incorporates orbital maneuvering capabilities, allowing reconfiguration to a more optimal configuration. Constellation reconfigurability is provided in two models depending on the nature of allowable orbital maneuvers, one incorporating only phasing maneuvers (changes in true anomaly) and one additionally allowing plane changes of inclination and/or right ascension of ascending node (RAAN).

\subsection{Satellite Agility} \label{subsec:agile}

The modeling of the concept of satellite agility is referred to as the \textit{agility model} and implements slewing about three principle axes $\hat{x}$, $\hat{y}$, and $\hat{z}$ with respect to a local-vertical local-horizontal reference frame in which the $\hat{z}$ points nadir, $\hat{x}$ points in the tangential direction of motion, and $\hat{y}$ results from a right-hand rule. The mission duration, $T_{\text{r}}$, is discretized for visibility and propagation of satellite states by the time step size $\Delta t > 0$, resulting in the number of time steps, $T = T_{\text{r}}/\Delta t$, where the set of time steps is given as $\mathcal{T} = \{1, 2, \ldots, T\}$.

The slewing about each axis $\hat{x}, \hat{y}$, and $\hat{z}$, as controlled by the Euler angle decision variables $\alpha_{\tau},~\beta_{\tau}$, and $\gamma_{\tau}$ about each axis, respectively, occurs at equally spaced attitude control opportunities with a step size of $\Delta \tau \ge \Delta t > 0$. Furthermore, the attitude control opportunities are contained within $\mathcal{T}^\prime = \{1, 2, \ldots, T_{\text{r}}/\Delta \tau\}$, which is a subset of $\mathcal{T}$ with $T_{\text{r}}/\Delta \tau$ total control opportunities. Attitude control is performed to provide coverage of a set of $P$ total targets $\mathcal{P}=\{1, 2, \ldots, P\}$ in which each individual target is denoted as $p \in \mathcal{P}$ with associated geodetic coordinates. The objective of the agility model is to minimize the angular difference, $\theta_{{\tau}p}$, between the current pointing direction $\bm{D}_{\tau}$ and the target pointing direction $\bm{T}_{{\tau}p}$ to target $p$ at control opportunity ${\tau} \in \mathcal{T}^\prime$, defined as:
\begin{equation}
    \theta_{{\tau}p} = \arccos{\left( \frac{\bm{D}_{\tau} \cdot \bm{T}_{{\tau}p}}{||\bm{D}_{\tau}|| \ ||\bm{T}_{{\tau}p}||} \right)}
\end{equation}
where $\bm{T}_{{\tau}p}$ is the vector originating from the center of the satellite body position and pointing to target $p$ at control opportunity $\tau$:
\begin{equation}
    \bm{T}_{{\tau}p} = \frac{\bm{P}_{\tau p} - \bm{r}_{\tau}}{|| \bm{P}_{\tau p} - \bm{r}_{\tau} ||}
\end{equation}
and where the satellite states, those being position $\bm{r}_{\tau}$ and velocity, are obtained via orbit propagation, and $\bm{P}_{\tau p}$ is the target position. The specific propagation method utilized in this paper is noted in Sec.~\ref{subsec:sim_params}.

\begin{figure}[!ht]
    \centering
    \includegraphics[width=0.33\textwidth]{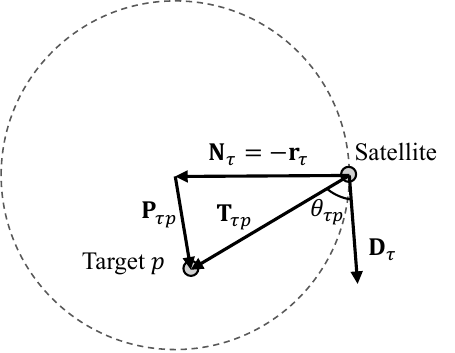}
    \caption{Visualization of agility model vector parameters.}
    \label{fig:Agile_Parameters}
\end{figure}

The minimization of the angular difference maximizes visibility through optimal pointing directions. The current pointing direction, $\bm{D}_{\tau}$, is computed at control opportunity ${\tau}$ via: 
\begin{equation}
    \bm{D}_{\tau} = M(\alpha_{\tau}, \beta_{\tau}, \gamma_{\tau}) \bm{N}_{\tau} 
\end{equation}
where $\bm{N}_{\tau} = -\bm{r}_{\tau}$ is the current nadir pointing direction at control opportunity ${\tau}$ and $M$ is a rotation matrix calculated in Appendix~A. Each pointing direction, those being $\bm{D}_{\tau}$, $\bm{T}_{{\tau}p}$, and $\bm{N}_{\tau}$, is defined with the origin at the satellite center. Figure~\ref{fig:Agile_Parameters} provides a visualization of the vector parameters used in the agility model.

The optimal slewing angles at each control opportunity are obtained through the optimization formulation shown in Formulation~\eqref{eq:agile}, which may be performed individually for a number of satellites:
\begin{subequations}
    \begin{alignat}{2}
\min \quad & \sum_{p \in \mathcal{P}} \sum_{{\tau} \in \mathcal{T}^\prime} \theta_{{\tau}p} \label{eq:agile_obj} && \\
\text{s.t.} \quad & |\alpha_{\tau} - \alpha_{{\tau}-1}| \leq \dot{\alpha} \Delta {\tau}, && \forall {\tau}\in \mathcal{T}^\prime \label{eq:agile_alpha_dot}\\ 
& |\beta_{\tau} - \beta_{{\tau}-1}| \leq \dot{\beta} \Delta {\tau}, && \forall {\tau}\in \mathcal{T}^\prime \label{eq:agile_beta_dot}\\ 
& |\gamma_{\tau} - \gamma_{{\tau}-1}| \leq \dot{\gamma} \Delta {\tau}, && \forall {\tau}\in \mathcal{T}^\prime \label{eq:agile_gamma_dot}\\
& \alpha_{\tau},~\beta_{\tau},~\gamma_{\tau} \in [-\zeta, \zeta], \qquad && \forall {\tau} \in \mathcal{T}^\prime \label{eq:agile_euler_limits}
    \end{alignat}
    \label{eq:agile}
\end{subequations}

Objective function~\eqref{eq:agile_obj} minimizes the sum of angular differences for all targets at all control opportunities, thus optimizing the slewing of a given satellite. Constraints~\eqref{eq:agile_alpha_dot}, \eqref{eq:agile_beta_dot}, and \eqref{eq:agile_gamma_dot} restrict the slewing rate of the satellite to a maximum of $\dot{\alpha}$, $\dot{\beta}$, and $\dot{\gamma}$, respectively, represented as the difference between the Euler angle at control opportunity ${\tau}-1$ to control opportunity ${\tau}$ being less than or equal to the maximum slewing rate multiplied by the time between control opportunities. It is assumed that all slewing angles prior to the first control opportunity are set to zero for nadir pointing such that $\alpha_0=\beta_0=\gamma_0=0$. Constraints~\eqref{eq:agile_euler_limits} limit the maximum slewing angle in all directions to between $-\zeta$ and $\zeta$. While this formulation assumes that the maximum slewing angle in all directions is identical, it should be noted that each maximum angle may be defined differently. 

The optimal slewing angles $\alpha^{\ast}_{\tau},~\beta^{\ast}_{\tau}$, and $\gamma^{\ast}_{\tau}$ at control opportunity $\tau \in \mathcal{T}^\prime$, obtained from Formulation~\eqref{eq:agile}, are applied to the propagated satellite states for the corresponding time steps in the range of $1 + (\tau - 1) \left( \Delta \tau / \Delta t \right) \le t \le \tau \left( \Delta \tau / \Delta t \right)$, which is repeated for all $\tau \in \mathcal{T}^\prime$. At each time step within this interval, the visibility $V_t = 1$ if any target $p \in \mathcal{P}$ is within a provided conical FOV, and $V_t = 0$ otherwise. The observation reward obtained is then calculated through the degradation caused by the change in look angle. If the look angle is nadir-directional, the reward is one, while if the look angle is not, the reward is degraded according to Eq.~\eqref{eq:Degredation}, derived from Ref.~\cite{CHATTERJEE2022}: 
\begin{equation}
    z = \sum_{t \in \mathcal{T}} V_t \left( 1 - \frac{|\alpha_{\tau}|+|\beta_{\tau}|+|\gamma_{\tau}|}{2(\alpha_{\max}+\beta_{\max}+\gamma_{\max})} \right)
    \label{eq:Degredation}
\end{equation}
where $z$ is the sum of all degraded observation rewards. Additionally, $\alpha_{\max},~\beta_{\max}$, and $\gamma_{\max}$ are equal to $\zeta$ as defined in constraint~\eqref{eq:agile_euler_limits}, a limitation in place additionally from Ref.~\cite{CHATTERJEE2022}.

All parameters, sets, and decision variables of the agility model are listed in Table~\ref{tab:agile_params}.

\begin{table}[!ht]
    \centering
    \caption{Agility model parameters, sets, and decision variables.}
    \begin{tabular}{ l l }
         \hline \hline
         Symbol & Definition \\
         \hline
         \multicolumn{2}{l}{Parameters} \\
         $\Delta {\tau} \ge \Delta t$ & Time step size between control opportunities \\
         $\Delta t > 0$ & Time step size between discrete time steps \\
         $T_{\text{r}} \ge 0$ & Mission duration \\
         $T$ & Total number of time steps \\
         $P > 0$               & Total number of targets \\
         $\theta_{{\tau}p}$ & Angular difference between $\bm{D}_{\tau}$ and $\bm{T}_{{\tau}p}$ for target $p \in \mathcal{P}$ at control opportunity ${\tau} \in \mathcal{T}^\prime$ \\
         $\bm{D}_{\tau}$ & Satellite pointing direction at control opportunity ${\tau} \in \mathcal{T}^\prime$ \\
         $\bm{T}_{{\tau}p}$ & Pointing direction for target $p \in \mathcal{P}$ at control opportunity ${\tau} \in \mathcal{T}^\prime$ \\
         $\bm{r}_{\tau}$ & Satellite position vector at control opportunity $\tau \in \mathcal{T}^\prime$\\
         $\bm{P}_{\tau p}$ & Position vector of target $p \in \mathcal{P}$ at control opportunity $\tau \in \mathcal{T}^\prime$ \\
         $M$ & Rotation matrix subject to $\alpha_{\tau}$, $\beta_{\tau}$, and $\gamma_{\tau}$ at control opportunity ${\tau} \in \mathcal{T}^\prime$ \\
         $\bm{N}_{\tau}$ & Satellite nadir pointing direction at control opportunity ${\tau} \in \mathcal{T}^\prime$ \\
         $\dot{\alpha} \ge 0$    & Maximum slewing rate about the $\hat{x}$ axis \\
         $\dot{\beta} \ge 0$     & Maximum slewing rate about the $\hat{y}$ axis \\
         $\dot{\gamma} \ge 0$    & Maximum slewing rate about the $\hat{z}$ axis \\
         $\zeta \ge 0$           & Maximum slewing angle about all axes \\
         $V_t$             & $\begin{cases}
                                  1, & \text{if target $p \in \mathcal{P}$ is visible at time step $t \in \mathcal{T}$ with slewing applied} \\
                                  0, & \text{otherwise}
                              \end{cases}$ \\
         \multicolumn{2}{l}{Sets} \\
         $\mathcal{T}$ & Set of time steps (index $t$, cardinality $T$) \\
         $\mathcal{T}^\prime \subseteq \mathcal{T}$ & Set of control opportunities (index ${\tau}$, cardinality $T_{\text{r}}/\Delta \tau$) \\
         $\mathcal{P}$ & Set of targets (index $p$, cardinality $P$) \\
         \multicolumn{2}{l}{Decision variables} \\
         $\alpha_{\tau}$ & Euler angle about the $\hat{x}$ axis at control opportunity ${\tau} \in \mathcal{T}^\prime$ \\
         $\beta_{\tau}$  & Euler angle about the $\hat{y}$ axis at control opportunity ${\tau} \in \mathcal{T}^\prime$ \\
         $\gamma_{\tau}$ & Euler angle about the $\hat{z}$ axis at control opportunity ${\tau} \in \mathcal{T}^\prime$ \\
         \hline \hline
    \end{tabular}
    \label{tab:agile_params}
\end{table}

\subsection{Constellation Reconfigurability} \label{subsec:recon}

The concept constellation reconfigurability implements the capability of satellite orbital maneuverability between various orbital slots over a given number of equally-spaced stages. The modeling of this CONOPS utilizes nadir-directional satellites with a conical FOV, additionally allowing satellites within the constellation to perform maneuvers and reconfiguration of the constellation according to the Multistage Constellation Reconfiguration Problem (MCRP) formulation, as proposed in Refs.~\cite{Lee2022,Lee2023,lee2024deterministic}. The objective of the MCRP is to maximize the observation rewards gathered by a constellation via optimal orbital maneuvers throughout a given number of stages for reconfiguration and several destination orbital slots for transfer. Constellation reconfigurability in this paper consists of two different models: the first allows only phasing maneuvers, while the second additionally permits changes in inclination and/or RAAN. Comparing the performances of these models allows us to characterize the impact of different orbital maneuvers on observation performance.

The MCRP formulation includes various sets that contain mission characteristics. The mission duration is defined as $T_{\text{r}}$, and the number of discrete time steps during the mission is defined as $T = T_{\text{r}}/\Delta t$ with time step size $\Delta t > 0$, where all time steps are contained in the set $\mathcal{T} = \{1, 2, \ldots, T\}$. Additionally, the set of stages $\mathcal{S} = \{0, 1, 2, \ldots, S\}$ contains the stages from zero (the initial stage of the constellation before reconfiguration) to the total number of stages $S$ at which the constellation can reconfigure. The number of time steps and the number of stages then define the set of stage time steps, $\mathcal{T}_{\text{s}}=\{1,2,\ldots, T_{\text{s}}\}$, where $T_{\text{s}}=T/S$ is the number of time steps in each stage $s\in \mathcal{S} \setminus \{0\}$. The set of satellites is defined as $\mathcal{K}=\{1,2,\ldots, K\}$, containing $K$ total satellites and their associated set of orbital slots at each stage, $\mathcal{J}_s^k=\{1,2,\ldots, J_s^k\}$, where $J_s^k$ is the total number of orbital slots for satellite $k\in \mathcal{K}$ in stage $s$. While the MCRP in Refs.~\cite{Lee2022,Lee2023,lee2024deterministic} considers a heterogeneous set of satellites in which some satellites are capable of maneuvers and others are not, this paper assumes a homogeneous set of satellites in which all satellites are capable of maneuvers. The final set is the set of target points, $\mathcal{P}=\{1,2,\ldots, P\}$, where $P$ is the total number of targets. 

Two binary variables are utilized by the MCRP formulation to control the orbital maneuvers of satellites or indicate the visibility of a target, respectively. The first is the binary decision variable $x_{ij}^{sk}$ that controls the transfer of orbital slots for satellite $k$ during stage $s$, which is defined as one in the event that the satellite transfers from orbital slot $i\in \mathcal{J}_{s-1}^k$ of stage $s-1$ to orbital slot $j\in \mathcal{J}_s^k$ of stage $s$, and defined as zero otherwise. The second is the binary indicator variable $y_{tp}^s$, which is defined as one in the event that target $p\in \mathcal{P}$ is visible to a required number of satellites during stage $s$ at time step $t \in \mathcal{T}_{\text{s}}$, and defined as zero otherwise.

Finally, the MCRP formulation also contains various parameters. The first parameter is the cost $c_{ij}^{sk} \ge 0$ of transfer from one orbital slot $i\in \mathcal{J}_{s-1}^k$ to the next orbital slot $j\in \mathcal{J}_s^k$, and the associated maximum transfer cost allotted for the mission, $c_{\max}^k \ge 0$, for satellite $k$. The next parameter is the observation reward, $\pi_{tp}^s \ge 0$, available for target $p$, in stage $s$, and at time step $t$, which is sought to be obtained throughout the mission duration. The observation rewards are subject to a coverage requirement $r_{tp}^s \in \mathbb{N}$, also for target $p$ at time step $t$ in stage $s$, defined as the number of satellites required to gain the associated reward. The final parameter is the binary VTW, $V_{tjp}^{sk}$, which is defined as one in the event that target $p$ is in view of satellite $k$ in orbital slot $j$ in stage $s$ at time step $t$ via the conical FOV provided to the satellite, and defined as zero if not. The parameter $V_{tjp}^{sk}$ of every target $p \in \mathcal{P}$ is obtained through the propagation of orbital slot $j$ of satellite $k$ for all times $t \in \mathcal{T}_{\text{s}}$ and all stages $s \in \mathcal{S} \setminus \{0\}$, and the specific propagation method utilized in this paper is noted in Sec.~\ref{subsec:sim_params}.

The mathematical formulation of the MCRP is shown as follows in Formulation~\eqref{eq:mcrp}~\cite{Lee2022,Lee2023,lee2024deterministic}:
\begin{subequations}
    \begin{alignat}{2}
\max \quad & z = \sum_{s\in\mathcal{S}\setminus\{0\}} \sum_{t\in\mathcal{T}_{\text{s}}} \sum_{p\in\mathcal{P}} \pi_{tp}^s y_{tp}^s \label{eq:mcrp_obj} \\
\text{s.t.} \quad & \sum_{j\in\mathcal{J}_{1}^k} x_{ij}^{1k} = 1, \quad && \forall k\in\mathcal{K}, i\in\mathcal{J}_0^k \label{eq:mcrp_init} \\
& \sum_{j\in\mathcal{J}_{s+1}^k}x_{ij}^{s+1,k}-\sum_{q\in\mathcal{J}_{s-1}^k}x_{qi}^{sk} = 0, && \forall s\in\mathcal{S}\setminus\{0,S\}, \forall k\in\mathcal{K}, \forall i\in\mathcal{J}_{s}^k \label{eq:mcrp_flow} \\
& \sum_{k\in\mathcal{K}} \sum_{i\in\mathcal{J}_{s-1}^k} \sum_{j\in\mathcal{J}_{s}^k} V_{tjp}^{sk} x_{ij}^{sk} \ge r_{tp}^s y_{tp}^s, \qquad && \forall s \in\mathcal{S}\setminus\{0\}, \forall t\in\mathcal{T}_{\text{s}}, \forall p\in\mathcal{P} \label{eq:mcrp_cov} \\
& \sum_{s\in\mathcal{S}\setminus\{0\}} \sum_{i\in\mathcal{J}_{s-1}^k} \sum_{j\in\mathcal{J}_{s}^k} c_{ij}^{sk} x_{ij}^{sk} \le c_{\max}^{k}, && \forall k \in\mathcal{K} \label{eq:mcrp_cost} \\
& x_{ij}^{sk} \in \{0,1\}, && \forall s\in\mathcal{S}\setminus\{0\}, \forall k\in\mathcal{K}, \forall i\in\mathcal{J}_{s-1}^k, \forall j\in\mathcal{J}_{s}^k \label{eq:mcrp_x} \\
& y_{tp}^s \in \{0,1\}, && \forall s\in\mathcal{S}\setminus\{0\}, \forall t\in\mathcal{T}_{\text{s}}, \forall p\in\mathcal{P} \label{eq:mcrp_y}
    \end{alignat}
    \label{eq:mcrp}
\end{subequations}

The MCRP returns the optimal sequence of orbital transfer maneuvers that maximizes the observation reward [objective function~\eqref{eq:mcrp_obj}]. Constraints~\eqref{eq:mcrp_init} control the transfer for each satellite of the orbital slots from the initial conditions of stage zero, $\mathcal{J}_0^k$, such that only one orbital slot $j\in \mathcal{J}_1^k$ is selected for transfer in the first stage. Constraints~\eqref{eq:mcrp_flow} similarly control the transfer of the orbital slots for subsequent stages $s \in \mathcal{S} \setminus \{0, S\}$ to ensure that transfers from $i\in \mathcal{J}_s^k$ to $j\in \mathcal{J}_{s+1}^k$ can only occur if the previous transfer from $q \in \mathcal{J}_{s-1}^k$ resulted in arrival to orbital slot $i$. Constraints~\eqref{eq:mcrp_cov} link the coverage requirement for the definition of $y_{tp}^s$, ensuring that target $p$ is covered if and only if at least $r_{tp}^s$ satellites can view the target with consideration of the VTW, $V_{tjp}^{sk}$. Constraints~\eqref{eq:mcrp_cost} apply the cost $c_{ij}^{sk}$ to the transfers such that the maximum cost $c_{\max}^k$ is not exceeded. Finally, constraints~\eqref{eq:mcrp_x} and \eqref{eq:mcrp_y} define the domain of the decision and indicator variable, respectively. 

All parameters, sets, and decision/indicator variables of the MCRP are listed in Table~\ref{tab:recon_params}.

\begin{table}[!ht]
    \centering
    \caption{MCRP parameters, sets, and decision/indicator variables.}
    \resizebox{\textwidth}{!}{
    \begin{tabular}{ l l }
         \hline \hline 
         Symbol & Definition \\
         \hline 
         \multicolumn{2}{l}{Parameters} \\
         $T_{\text{r}} \ge 0$ & Mission duration \\
         $\Delta t > 0$ & Time step size between discrete time steps \\
         $T$ & Number of time steps \\
         $S \ge 0$ & Number of stages for reconfiguration \\
         $T_{\text{s}}$ & Number of time steps in each stage \\
         $K \ge 0$ & Number of satellites \\
         $J_s^k \ge 0$ & Number of orbital slots available to satellite $k \in \mathcal{K}$ in stage $s \in \mathcal{S} \setminus \{0\}$ \\
         $P > 0$ & Total number of targets \\
         $c_{ij}^{sk} \ge 0$  & Cost for satellite $k$ to transfer from orbital slot $i \in \mathcal{J}_{s-1}^k$ to orbital slot $j \in \mathcal{J}_s^k$ \\
         $c^k_{\max} \ge 0$   & Maximum transfer cost for satellite $k$ \\
         $\pi_{tp}^s \ge 0$   & Observation reward available for target $p \in \mathcal{P}$ at time step $t \in \mathcal{T}_{\text{s}}$ in stage $s \in \mathcal{S} \setminus \{0\}$ \\
         $r_{tp}^s \in \mathbb{N}$     & Coverage requirement for target $p \in \mathcal{P}$ at time step $t \in \mathcal{T}_{\text{s}}$ in stage $s \in \mathcal{S} \setminus \{0\}$ \\
         $V_{tjp}^{sk}$ & $\begin{cases}
                               1, & \text{if target $p \in \mathcal{P}$ is visible to satellite $k \in \mathcal{K}$ in orbital slot $j \in \mathcal{J}_s^k$ at time step $t \in \mathcal{T}_{\text{s}}$ in stage $s \in \mathcal{S} \setminus \{0\}$} \\
                               0, & \text{otherwise}
                           \end{cases}$\\
         \multicolumn{2}{l}{Sets} \\
         $\mathcal{T}$            & Set of time steps (index $t$, cardinality $T$) \\
         $\mathcal{S}$            & Set of stages (index $s$, cardinality $S$) \\
         $\mathcal{T}_{\text{s}}$ & Set of time steps in each stage (index $t$, cardinality $T_{\text{s}}$)\\
         $\mathcal{K}$            & Set of satellites (index $k$, cardinality $K$) \\
         $\mathcal{J}_s^k$        & Set of orbital slots (indices $i,~j$, cardinality $J_s^k$)\\
         $\mathcal{P}$ & Set of targets (index $p$, cardinality $P$) \\
         \multicolumn{2}{l}{Decision and indicator variables} \\
         $x_{ij}^{sk}$ & $\begin{cases}
                              1, & \text{if satellite $k \in \mathcal{K}$ transfers from orbital slot $i \in \mathcal{J}_{s-1}^k$ to orbital slot $j \in \mathcal{J}_s^k$} \\
                              0, & \text{otherwise}
                          \end{cases}$\\
         $y_{tp}^s$    & $\begin{cases}
                              1, & \text{if target $p \in \mathcal{P}$ is visible to any satellite in orbital slot $j \in \mathcal{J}_s^k$ at time step $t \in \mathcal{T}_{\text{s}}$ in stage $s \in \mathcal{S} \setminus \{0\}$} \\
                              0, & \text{otherwise}
                          \end{cases}$\\
         \hline \hline
    \end{tabular}
    }
    \label{tab:recon_params}
\end{table}

\subsubsection{Phasing-Restricted Reconfigurability} \label{subsec:phasing_model}

The first application of the MCRP is restricted to phasing maneuvers, only incorporating candidate orbital slots that vary in phase (true anomaly or argument of latitude, dependent upon eccentricity). This variant of the MCRP is referred to as the \textit{phasing-restricted reconfigurability model}. High-thrust impulsive maneuver cost calculations for phasing changes are performed using the phasing rendezvous algorithm from Ref.~\cite{Vallado2013} to define values of $c_{ij}^{sk}$. The phasing rendezvous algorithm implemented allows up to four complete revolutions of the target orbital slot and of the transfer orbit utilized to perform the rendezvous.

\subsubsection{Unrestricted Reconfigurability} \label{subsec:plane_change_model}

The second application of the MCRP incorporates orbital slots that enable both plane changes and phasing maneuvers (varying in inclination, RAAN, and phase). This variant of the MCRP is referred to as the \textit{unrestricted reconfigurability} model. The additional maneuverability provided through changes in the orbital plane alongside phase changes is an additional comparison to demonstrate the extent of the capabilities within constellation reconfigurability. High-thrust impulsive maneuver cost calculations for changes in the orbital plane are performed using analytical algorithms from Ref.~\cite{Vallado2013}. These algorithms are for the one or two-impulse direct transfer cost from one orbit to another. The possible transfers include 1) phasing only, 2) inclination change only, 3) RAAN change only, 4) simultaneous inclination and RAAN change, 5) inclination change followed by phasing, 6) RAAN change followed by phasing, and 7) simultaneous inclination and RAAN change followed by phasing. Any instance of phasing will occur after the plane change and in accordance with the phasing rendezvous algorithm previously mentioned. 

\section{Comparative Analysis} \label{sec:comparative_analysis}

A comprehensive comparative analysis of each CONOPS is conducted to determine the level of performance in response to TCs. The effectiveness is reported through the coverage of a given TC in terms of the sum of observation rewards, $z$. 

\subsection{Tropical Cyclone Data} \label{subsec:disaster_hist}

\begin{figure}[!ht]
    \centering
    \begin{subfigure}[h]{0.49\textwidth}
        \centering
        \includegraphics[width = \textwidth]{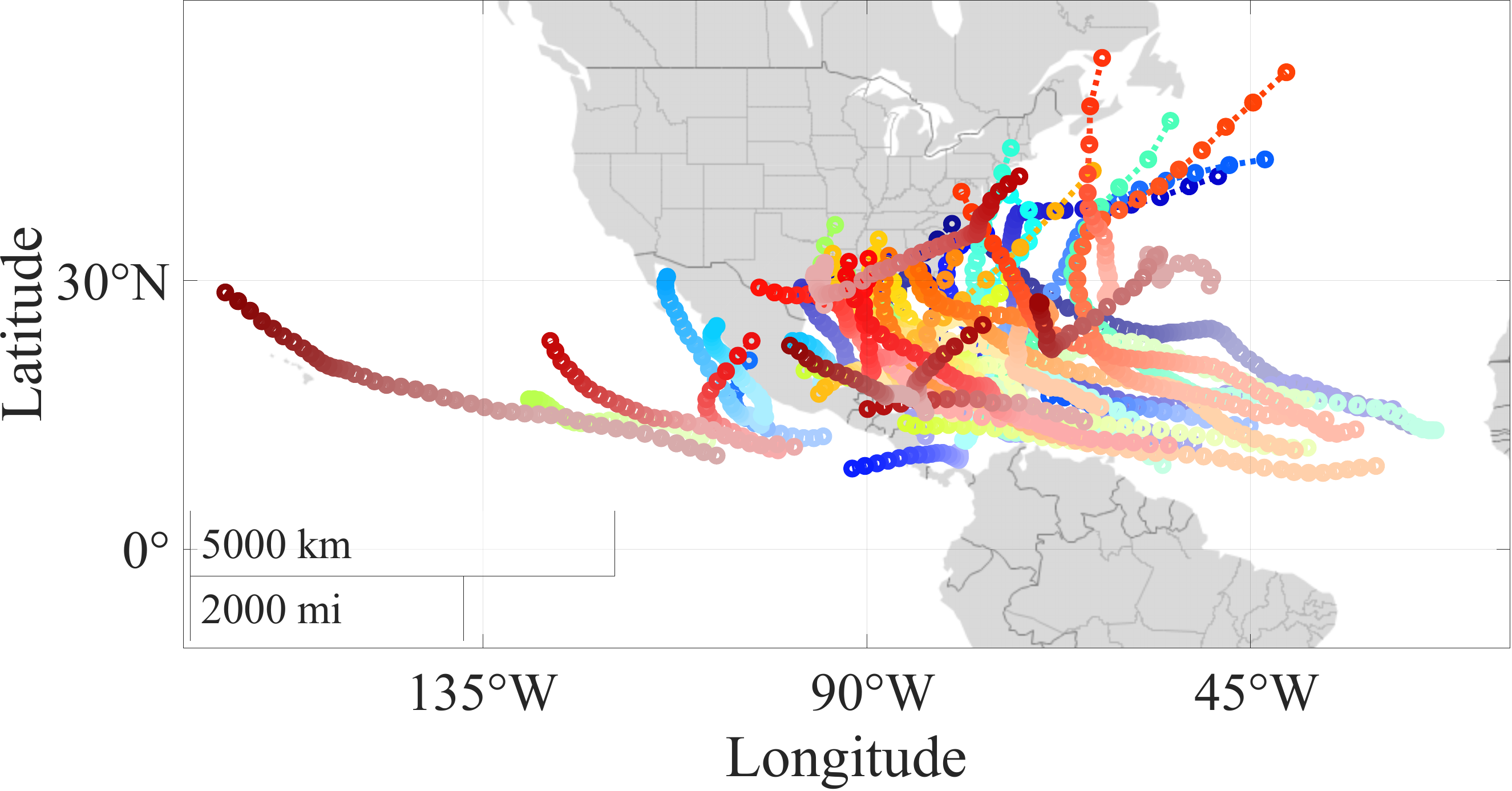}
        \caption{Western hemisphere.}
        \label{fig:Historical_Hurricanes}
    \end{subfigure}
    \hfill
    \begin{subfigure}[h]{0.49\textwidth}
        \centering
        \includegraphics[width = \textwidth]{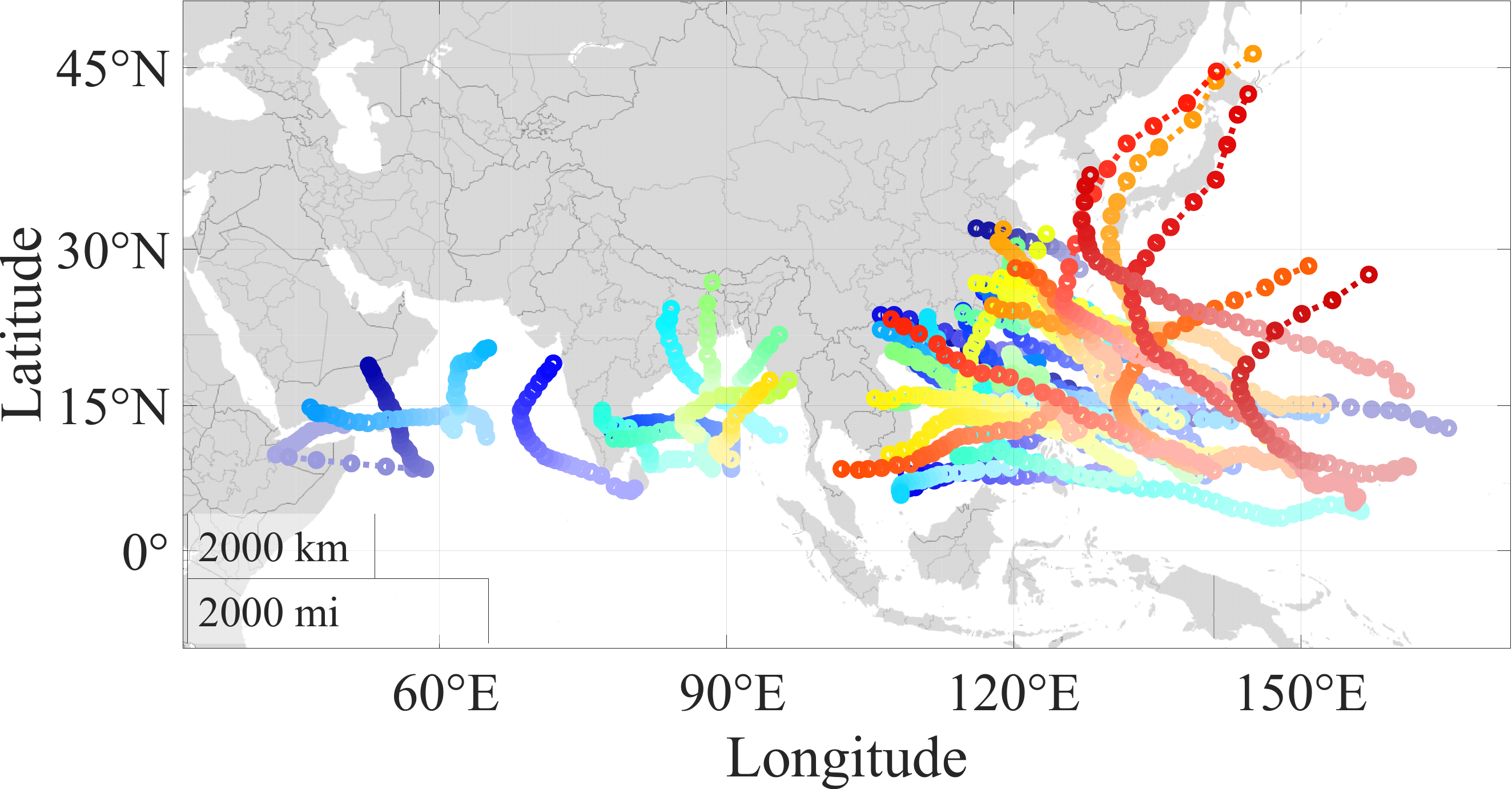}
        \caption{Eastern hemisphere.}
        \label{fig:Historical_Typhoons}
    \end{subfigure}
    \caption{Historical TC trajectories.}
    \label{fig:Historical_Cyclones}
\end{figure}

A data set of 100 historical TCs is used for evaluation. This highly populated data set provides a large variance in the length and duration of each TC. Out of the 100 TCs included in this data set, 80 have retired names, meaning that the impact of these storms is so severe that future uses of the same name are seen as inappropriate for sensitivity to those affected by the named storm \cite{RetiredNames}. Each TC is tracked along the latitude and longitude of the eye of the storm at six-hour intervals and the tracked data is retrieved from Ref.~\cite{WUnderground}. Additionally, each TC is tracked from the first occurrence of tropical storm status (winds between 39 and 73 mph \cite{TempestDefinitions_2022}) until the last occurrence of tropical storm status. All 100 TCs are depicted in Fig.~\ref{fig:Historical_Cyclones}, with hurricanes in Fig.~\ref{fig:Historical_Hurricanes} in the western hemisphere and typhoons/cyclones in Fig.~\ref{fig:Historical_Typhoons} in the eastern hemisphere. Each TC is modeled to be comprised of $P$ total points that exist for the entire duration, $T_{\text{r}}$, of the TC, but each point is only given observation rewards for its respective duration of six hours, thus only allowing a single point to contain observation rewards at any given time step and accounting for the motion of the TC.

\begin{figure}[!ht]
    \centering
    \includegraphics[width = 0.6\textwidth]{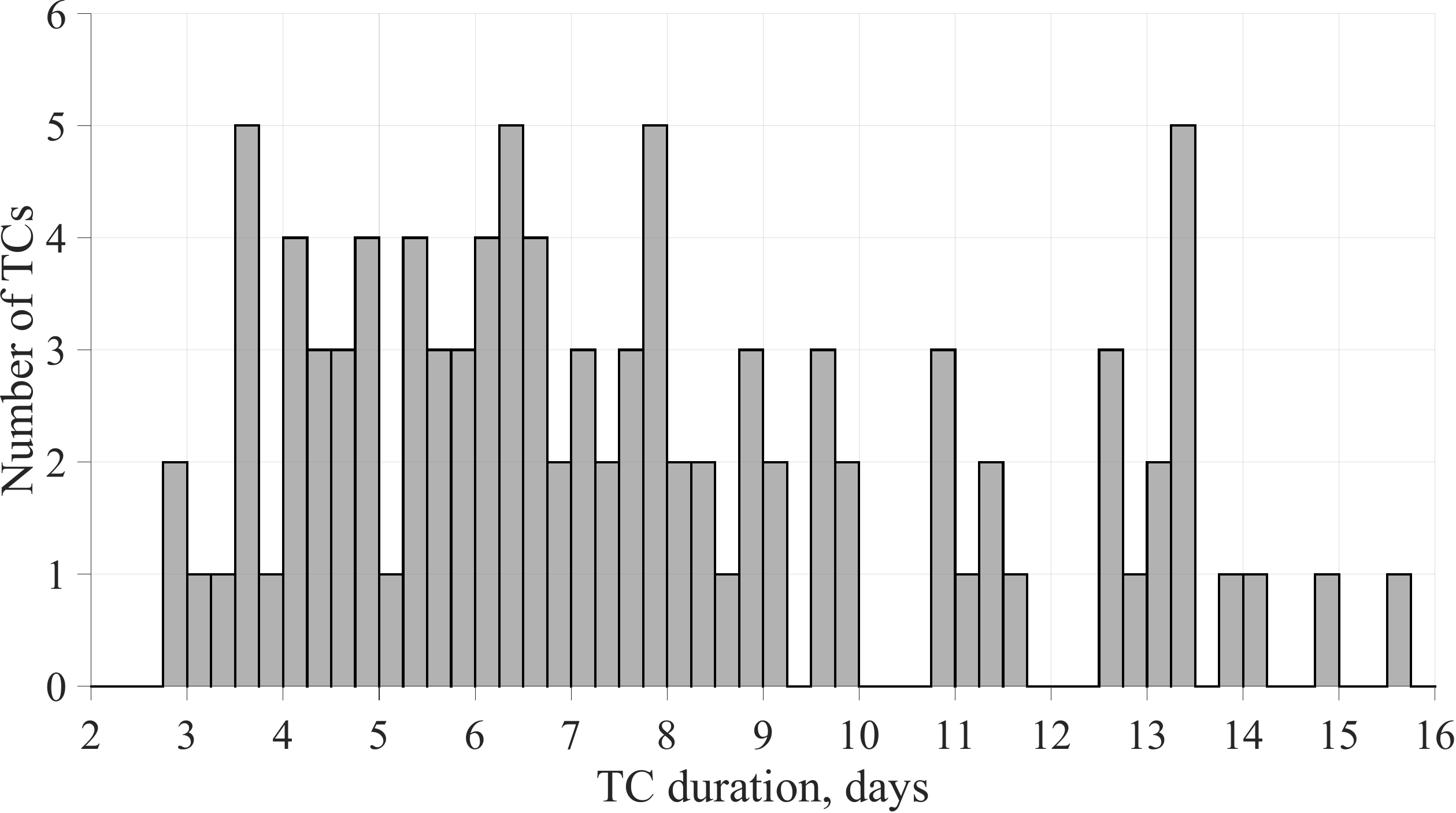}
    \caption{Histogram of the time duration of TCs.}
    \label{fig:TimeDist}
\end{figure}

The shortest TC is Cyclone Rumbia (2018) which lasted eight days, and the longest is Hurricane Florence (2018) which lasted 19 days. As a result of only tracking with tropical storm status rather than the entire disaster, $T_{\text{r}}$ for these TCs will be \SI{2.75}{days} and \SI{15.5}{days}, respectively. Additionally, Fig.~\ref{fig:TimeDist} shows a histogram of the varying duration of each TC, grouped by the same six-hour interval used for tracking.  

\subsection{CONOPS Parameters} \label{subsec:sim_params}

A variety of disaster monitoring satellite orbits, mainly those comprising the Disaster Monitoring Constellation (DMC), are used as a case study for the comparison of each CONOPS. Orbits from the DMC are chosen as they are designed for disaster monitoring \cite{Stephens2003DMC} and provide a large set of satellite orbits to choose from, varying in altitude, inclination, RAAN, and phase. Via Ref.~\cite{DMC_TLE}, the two-line element set for $28$ currently operating disaster monitoring satellites are available, of which five satellite orbits are selected based on proximity to a 7000-km semi-major axis, a metric chosen as a majority of the original $28$ satellites have a semi-major axis of $\SI{7000}{km} \pm \SI{100}{km}$. Only the subset of five satellites is chosen for consideration of computational runtime with respect to the 100 TCs used as data points. All classical orbital element (COE) data is true to an epoch of March 1st, 2023, at midnight (00:00:00) in Coordinated Universal Time (UTC), and the COEs for each chosen satellite are shown in Table~\ref{tab:DMC}. 

\begin{table}[!ht]
    \centering
    \caption{COEs for modeling, obtained via Ref.~\cite{DMC_TLE}.}
    \resizebox{\textwidth}{!}{
    \begin{threeparttable}
    \begin{tabular}{ l  r  r  r  r  r  r } 
        \hline \hline
        Satellite name & Semi-major axis, km & Eccentricity & Inclination, deg & RAAN, deg & Argument of periapsis, deg & True anomaly, deg \\
        \hline
        DMC 3-FM3    & 7006.01 & $\num{17.07e-4}$ & 97.72 & 307.83 & 77.52  & 104.88 \\
        DMC 3-FM1    & 6992.54 & $\num{8.03e-4}$  & 97.72 & 306.02 & 116.04 & 302.43 \\
        HUANJING 1B  & 7003.07 & $\num{48.93e-4}$ & 97.80 & 89.49  & 107.47 & 140.62 \\
        HUANJING 1A  & 7007.36 & $\num{39.24e-4}$ & 97.79 & 85.41  & 116.27 & 189.24 \\
        NIGERIASAT 1 & 6992.76 & $\num{41.58e-4}$ & 97.85 & 228.61 & 260.58 & 149.89 \\
        \hline \hline
    \end{tabular}
    \end{threeparttable}
    }
    \label{tab:DMC}
\end{table}

Each CONOPS is coded in MATLAB and leverages the functions of the Satellite Communications Toolbox \cite{MATLAB} for the satellite propagation and visibility profiles; the propagator used is the Simplified General Perturbation 4 model (SGP4). SGP4 accounts for orbital perturbations of near-Earth orbits (orbital period less than \SI{225}{minutes}) caused by the shape of the Earth, atmospheric drag, and solar radiation pressures \cite{MATLAB}. The visibility utilized in each model is a representation of whether or not the TC is within the conical FOV of a satellite in the constellation at the given time steps. Additionally, MATLAB's sequential convex programming algorithm via fmincon \cite{MATLAB} is used to solve the agility model [Formulation~\eqref{eq:agile}], while the commercial software package Gurobi Optimizer (version 11.0.0) is utilized in conjunction with YALMIP modeling \cite{YALMIP} to solve the MCRP [Formulation~\eqref{eq:mcrp}].

We define the parameters of each CONOPS as follows. The parameters assigned to all CONOPS include a conical FOV of $45$ degrees and a time step of $\Delta t = \SI{100}{seconds}$ between evaluations in propagation and visibility. The parameters assigned to the agility model include a maximum slewing rate $\dot{\alpha}=\dot{\beta}=\dot{\gamma}=\SI{3}{deg/s}$ as defined in Ref.~\cite{Lappas2002}, a maximum slewing angle $\zeta=\SI{35}{deg}$ as defined in Refs.~\cite{Chen2020,Karpenko2019}, and slewing opportunities occur every \SI{30}{minutes} ($\Delta {\tau}=\SI{1800}{seconds}$).

\begin{figure}[!ht]
    \centering
        \begin{subfigure}[h]{0.49\textwidth}
        \centering
        \includegraphics[width = 0.75\textwidth]{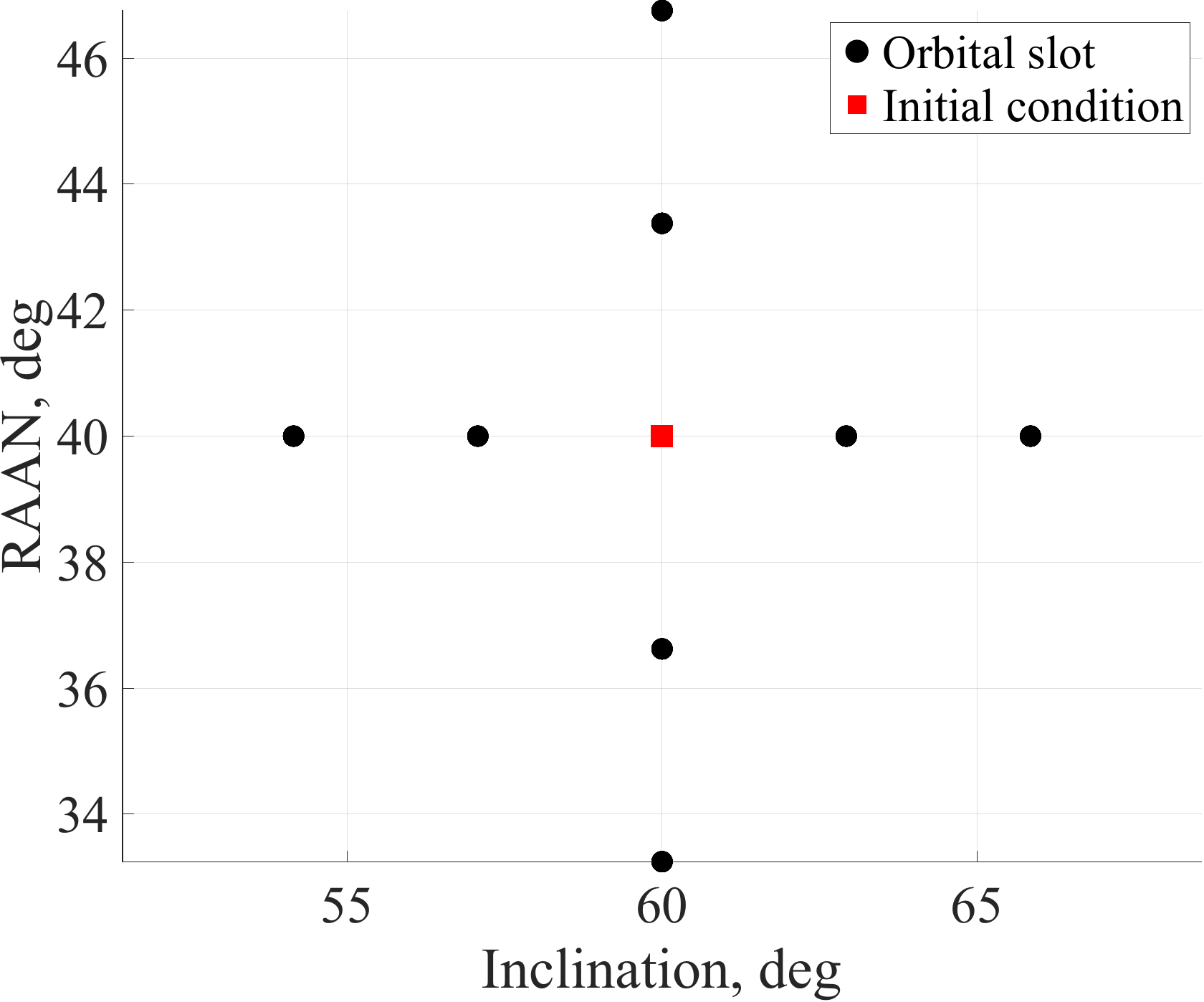}
        \caption{Planar orbital slots.}
        \label{fig:Planar}
    \end{subfigure}
    \begin{subfigure}[h]{0.49\textwidth}
        \centering
        \includegraphics[width = \textwidth]{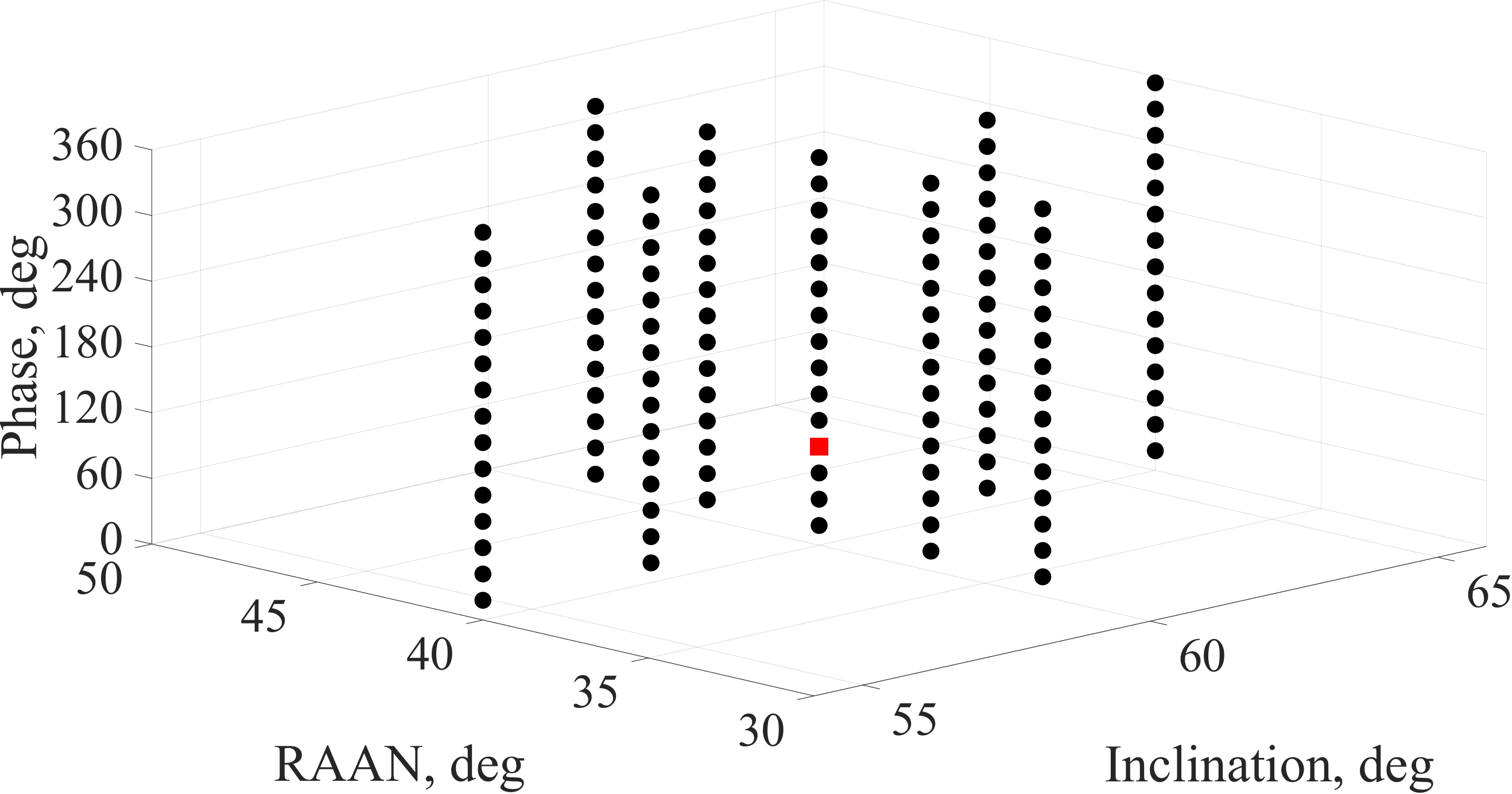}
        \caption{Three-dimensional orbital slots.}
        \label{fig:3D}
    \end{subfigure}
    \caption{Illustration of orbital slot options.}
    \label{fig:slot_options}
\end{figure}

The parameters assigned to both the phasing-restricted reconfigurability model and the unrestricted reconfigurability model include a requirement of only a single satellite to gain observations, set as $r_{tp}^s = 1, \forall p \in \mathcal{P}, \forall s \in \mathcal{S} \setminus \{0\}, \forall t \in \mathcal{T}_{\text{s}}$. To model the motion of each TC without reconfiguration stages, $\pi_{tp} = 1$ if $1 + (p-1)(T/P) \le t \le p(T/P), \forall p \in \mathcal{P}, \forall t \in \mathcal{T}$ and $\pi_{tp} = 0$ otherwise, which can be broken into $S$ stages by $\pi^s_{tp} = \pi_{t + (s-1)T_{\text{s}}, p}, \forall s \in \mathcal{S} \setminus \{0\}, \forall t \in \mathcal{T}_{\text{s}}, \forall p \in \mathcal{P}$. Additionally, the phasing-restricted reconfigurability model is assigned $J_s^k=10$ and $J_s^k=20$ equally spaced phasing orbital slots on the interval $[0, 360)$ degrees per satellite and stage with the inclusion of the initial phase and such that the options are identical from one stage to the next ($\mathcal{J}_{s-1}^k=\mathcal{J}_s^k,~\forall s\in \mathcal{S}\setminus\{0,1\}$). Similarly, the parameters assigned to the unrestricted reconfigurability model include orbital slots of equally spaced inclination, RAAN, and phase. The inclination and RAAN slots extend in the positive and negative directions of the initial slot with a shared slot at the initial conditions. All options ensure consideration of the budget, $c_{\max}^k$, such that the furthest options of inclination and RAAN would expend the entire budget as a single maneuver. The resultant total number of slot options is $J_s^k = \ell (2m-1)$, where $m$ is the number of planar options on either the inclination or RAAN axis, including the initial conditions, and $\ell$ is the number of phasing options. The orbital slots assigned for the comparative analysis include five inclination and RAAN options along each respective axis ($m=5$), with a shared option in the center, thus contributing $2m-1 = 9$ planar options. Additionally, $\ell=15$ phasing options are provided per plane, resulting in a total of $J_s^k=135$ orbital slots. Both reconfigurability models utilize a fuel budget of $c_{\max}^k=\SI{2}{km/s}, \forall k\in \mathcal{K}$. An illustrative visualization of the orbital slot options for the unrestricted reconfigurability model is provided in Fig.~\ref{fig:slot_options}, where the nine planar slots are depicted in Fig.~\ref{fig:Planar}, with the initial condition at the center, alongside all $135$ slots as a three-dimensional representation in Fig.~\ref{fig:3D}.

Due to the varying temporal scales of selected TCs, different numbers of stages are applied to the reconfigurability models to observe the effects of a change in the number of orbital slots and stages. Although the shortest TCs are only \SI{2.75}{days}, this duration is long enough to allow for more than a single stage to be realistic. Therefore, a combination of two and four stages is used, which allows a reconfiguration (at most frequent) every \SI{16.5}{hours}. 

As a result of the various combinations of parameters being applied to each CONOPS model, the abbreviations used for each combination are listed in Table~\ref{tab:abbreviations}.

\begin{table}[!ht]
    \centering
    \caption{Model abbreviations.}
    \begin{tabular}{ l l l }
        \hline \hline
         Model abbreviation & Corresponding model & Key parameters \\
         \hline
         Model~B  & Baseline model & - \\
         Model~A  & Agility model  & - \\
         Model~P1 & Phasing-restricted reconfigurability model & $S=2,~J=10$ \\
         Model~P2 & Phasing-restricted reconfigurability model & $S=2,~J=20$ \\
         Model~P3 & Phasing-restricted reconfigurability model & $S=4,~J=10$ \\
         Model~P4 & Phasing-restricted reconfigurability model & $S=4,~J=20$ \\
         Model~U1 & Unrestricted reconfigurability model       & $S=2,~J=135$ \\
         Model~U2 & Unrestricted reconfigurability model       & $S=4,~J=135$ \\
         \hline \hline
    \end{tabular}
    \label{tab:abbreviations}
\end{table}

\subsection{Results and Discussion} \label{subsec:results}

\begin{figure}[!ht]
    \centering
    \includegraphics[width = \textwidth]{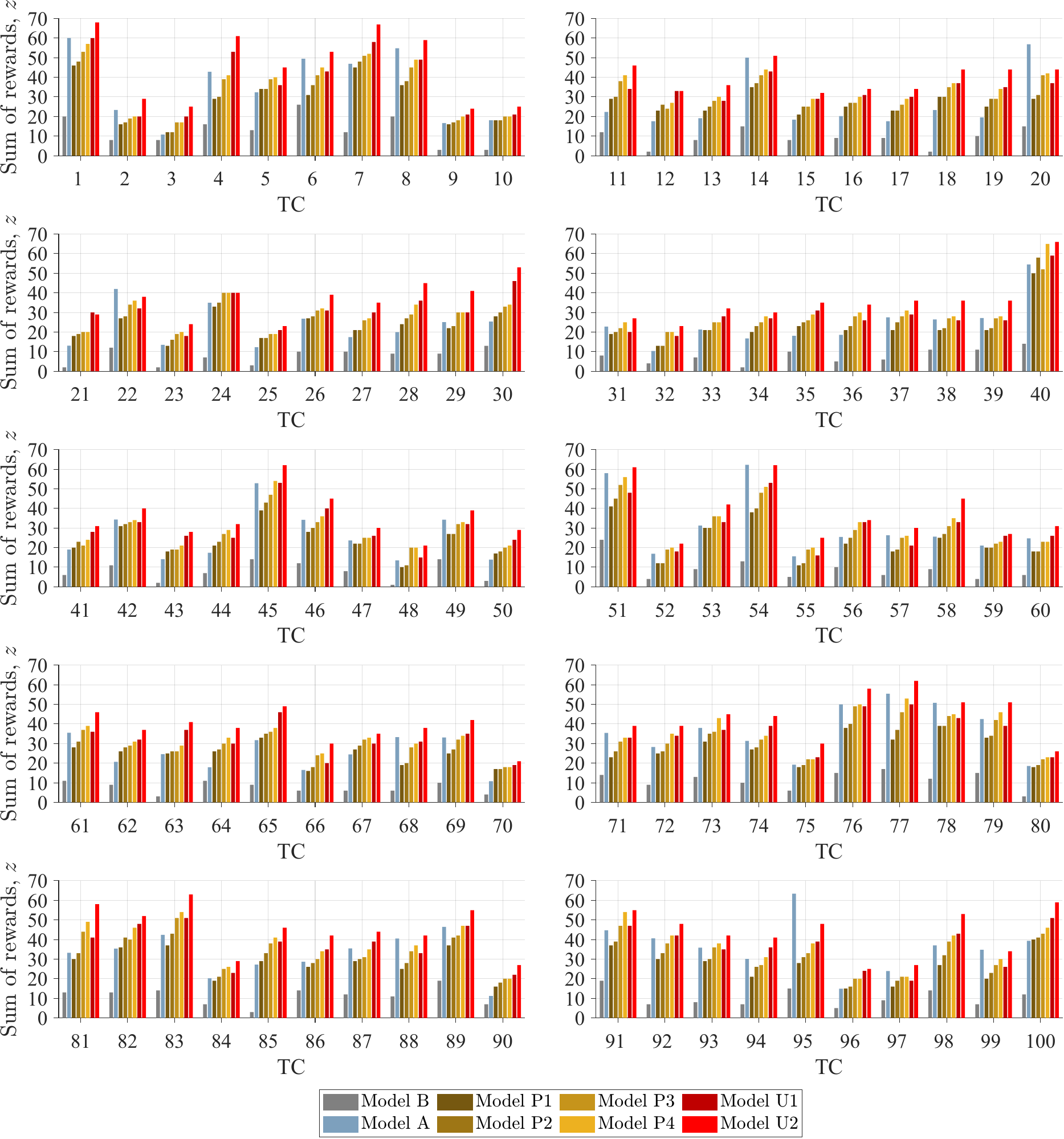}
    \caption{Sum of observation rewards, $z$, for all TCs.}
    \label{fig:Tempest_rewards}
\end{figure}

The results of simulating each CONOPS model subject to all 100 selected TCs are provided in this section. An additional set of results with an FOV of $30$ degrees is presented in Appendix~B. A summary of the sum of observation rewards, $z$, is provided in Fig.~\ref{fig:Tempest_rewards} with an individual grouping of all CONOPS models for each TC. Each group of bars is displayed, from left to right: Model~B in gray, Model~A in blue, Model~P1 through Model~P4 in varying shades of orange, and Model~U1 and Model~U2 in two shades of red. For all other figures displaying all CONOPS models, the same coloring scheme will apply.

\begin{figure}[!ht]
    \centering
    \includegraphics[width=0.6\linewidth]{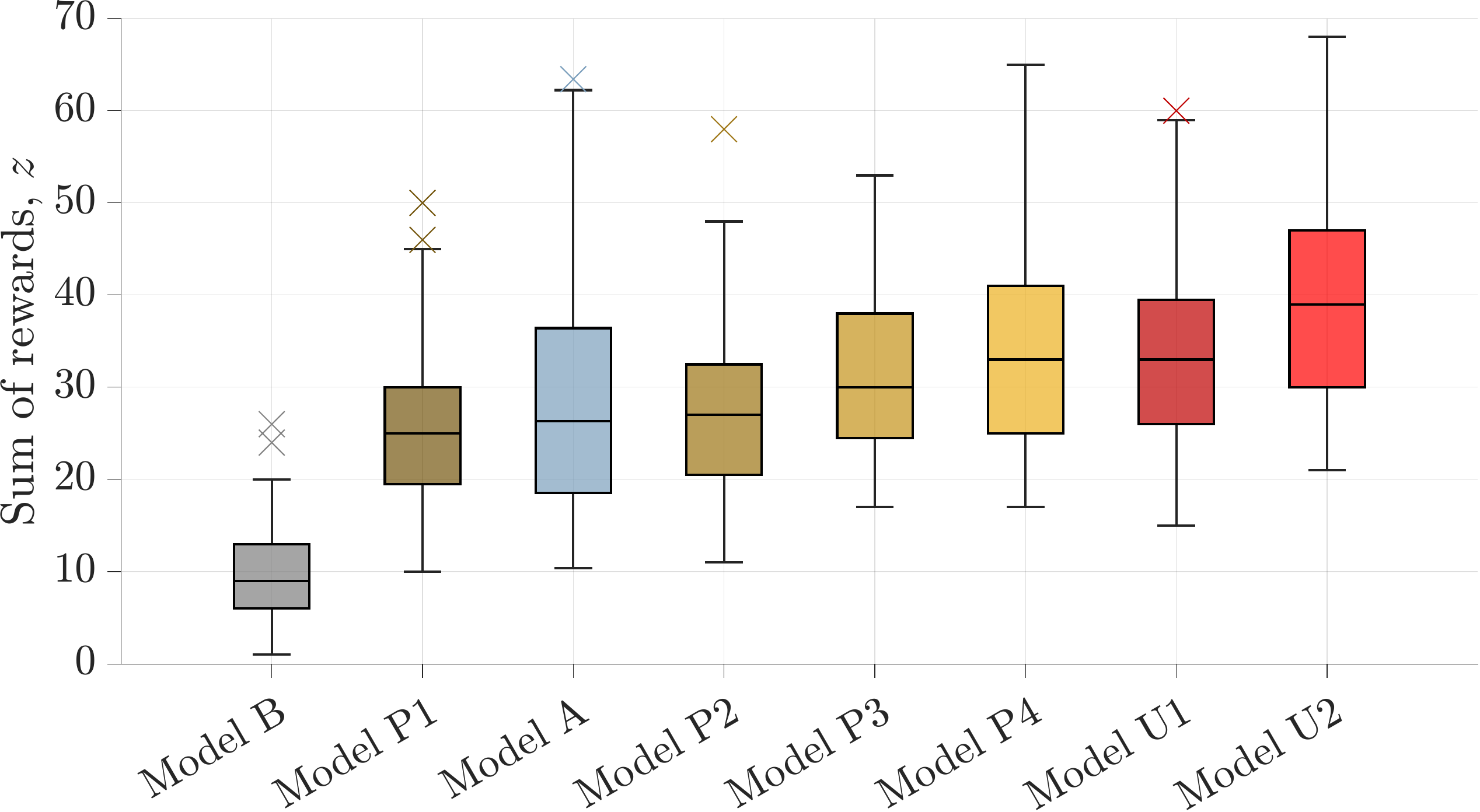}
    \caption{Sum of rewards organized by average performance.}
    \label{fig:box_Z}
\end{figure}

Additionally, Fig.~\ref{fig:box_Z} displays the sum of observation rewards, $z$, as a box chart organized from left to right in ascending order of the average increase in performance over Model~B. In the order that each model appears in Fig.~\ref{fig:box_Z}, Table~\ref{tab:Performance_over_baseline} provides statistics on the increase in performance of each model over Model~B, including the average, standard deviation, and ranges of the increase in performance. As reflected in Fig.~\ref{fig:box_Z} and Table~\ref{tab:Performance_over_baseline}, Model~P1 is the lowest-performing model on average with the lowest minimum increase and the second lowest maximum increase, Model~U2 is the highest-performing model on average with the highest minimum and maximum increase, and Model~A is the second lowest-performing model on average with the second lowest minimum increase and the lowest maximum performance.

\begin{table}[!ht]
    \centering
    \caption{Percent increase in performance of all models over Model~B.}
    \begin{tabular}{ l r r r }
         \hline \hline 
         Model & Average $\pm$ standard deviation, \% & Minimum, \% & Maximum, \% \\
         \hline 
         Model~P1 & $ 247.54 \pm 233.76 $ & 19.23  & 1400    \\
         Model~A  & $ 271.40 \pm 202.28 $ & 33.92  & 1245.05 \\
         Model~P2 & $ 273.55 \pm 251.76 $ & 38.46  & 1400    \\
         Model~P3 & $ 330.54 \pm 301.48 $ & 57.69  & 1900    \\
         Model~P4 & $ 359.64 \pm 316.41 $ & 73.08  & 1900    \\
         Model~U1 & $ 364.35 \pm 328.29 $ & 65.39  & 1900    \\
         Model~U2 & $ 450.92 \pm 369.92 $ & 103.85 & 2100    \\
         \hline \hline 
    \end{tabular}
    \label{tab:Performance_over_baseline}
\end{table}

In addition to Model~U2 being the highest-performing model in terms of the increase in performance over Model~B, Model~U2 outperforms all other models in a majority of (more than 90) TCs. The exact number of TCs in which each model outperforms the other models is organized in Table~\ref{tab:cross_reference_performance}; each numeric in the table corresponds to the number of TCs where the model in that column outperforms the model in the row. Table~\ref{tab:cross_reference_performance} reflects that while Model~U2 is the highest-performing on average, Model~A additionally outperforms Models~P1 and~P2 in more than half of the TCs, and Models P3, P4, and U1 in more than 20 of the TCs, as well as Model~U2 in five of the TCs. This indicates that Model~A has some cases in which it is the highest-performing model, despite not being the highest-performing on average. Model~U1 also outperforms Model~U2 in one TC, while no other models outperform Model~U2 in any TC. Finally, it should be noted that Models P3 and P4 outperform Model~U1 in 28 and 47 TCs, respectively, indicating that there are cases in which phasing-restricted reconfigurability may outperform reconfigurability additionally allowing plane changes.

\begin{table}[!ht]
    \centering
    \caption{The number of TCs in which the model of a column outperforms the model in the row.}
    \resizebox{\textwidth}{!}{
    \begin{tabular}{ l r r r r r r r r}
         \hline \hline
         & Model~B & Model~A & Model~P1 & Model~P2 & Model~P3 & Model~P4 & Model~U1 & Model~U2 \\
         \hline 
         Model~B  & -   & 100 & 100 & 100 & 100 & 100 & 100 & 100 \\
         Model~A  & 0   & -   & 35  & 45  & 65  & 77  & 73  & 95  \\
         Model~P1 & 0   & 65  & -   & 83  & 100 & 100 & 100 & 100 \\
         Model~P2 & 0   & 55  & 0   & -   & 91  & 100 & 98  & 100 \\
         Model~P3 & 0   & 35  & 0   & 4   & -   & 83  & 62  & 99  \\
         Model~P4 & 0   & 23  & 0   & 0   & 0   & -   & 42  & 99  \\
         Model~U1 & 0   & 27  & 0   & 0   & 28  & 47  & -   & 97  \\
         Model~U2 & 0   & 5   & 0   & 0   & 0   & 0   & 1   & -   \\
         \hline \hline
    \end{tabular}
    }
    \begin{tablenotes}
        \item[-] A hyphen (-) indicates that the models are the same, and thus cannot outperform itself.
    \end{tablenotes}
    \label{tab:cross_reference_performance}
\end{table}

Through the data presented above, some key observations are presented. The first observation drawn from the results is that constellation reconfigurability has the potential to outperform satellite agility in response to TCs when given adequate parameters. This is primarily shown through the average increase in performance of Model~U2 depicted in Fig.~\ref{fig:box_Z} and Table~\ref{tab:Performance_over_baseline}, where Model~U2 has the highest performance in the average, minimum, and maximum. Additionally, this observation is shown in Table~\ref{tab:cross_reference_performance}, where Model~U2 outperforms every other model in more than 95 out of 100 total TCs. Table~\ref{tab:ModelU2_performance} depicts the performance of Model~U2 above each other model, including the same information provided by Table~\ref{tab:cross_reference_performance}, further showing that Model~U2 outperforms all other models on average. However, Table~\ref{tab:ModelU2_performance} further reflects that there are cases in which Models~A and~U1 outperform Model~U2. Furthermore, the other reconfigurability models with higher numbers of orbital slots and stages, those being Models P3, P4, and U1, also outperform Models A, P1, and P2 in a majority of TCs, though not to the extent of Model~U2. This mainly results from the frequent reconfiguration opportunities and the high level of flexibility present within constellation reconfigurability, especially when plane changes are present as optional transfers. Additionally, the difference in parameters between the models that did and did not outperform Model~A are minor, suggesting that slight increases in either the number of stages or the number of orbital slots would increase the results even further.

\begin{table}[!ht]
    \centering
    \caption{Percent increase in performance of Model~U2 over other models.}
    \begin{tabular}{ l r r r }
         \hline \hline 
         Model & Average $\pm$ standard deviation, \% & Minimum, \% & Maximum, \% \\
         \hline 
         Model~A  & $48.52 \pm 37.35$ & -24.32 & 141.55 \\
         Model~P1 & $60.48 \pm 20.39$ & 21.21  & 127.27 \\
         Model~P2 & $49.30 \pm 18.34$ & 13.79  & 108.33 \\
         Model~P3 & $29.15 \pm 11.74$ & 0      & 60.61  \\
         Model~P4 & $20.56 \pm 10.92$ & 0      & 55.88  \\
         Model~U1 & $20.99 \pm 10.88$ & -3.33  & 56.25  \\
         \hline \hline 
    \end{tabular}
    \label{tab:ModelU2_performance}
\end{table}

A secondary observation drawn from the results is the importance of orbital slot flexibility in comparison to the number of stages available for reconfiguration. The results reflect that the flexibility and availability of more orbital slots are more important than the number of stages available for transfer, highlighted in Table~\ref{tab:cross_reference_performance} through the 62 and 42 TCs where Model~U1 outperformed Models~P3 and~P4, respectively, despite having fewer stages for reconfiguration. This observation is further reflected through the four TCs where Model~P2 outperformed Model~P3, where Model~P2 has more orbital slots but fewer stages than Model~P3. Furthermore, this observation is depicted in Table~\ref{tab:Performance_over_baseline} through the higher average performance of Model~U1 than Models P3 and P4, as well as the higher minimum increase in performance than Model~P3. This difference in performance despite having fewer stages is a direct result of the added flexibility in the orbital slots available, being expanded to plane changes in inclination and/or RAAN, as well as in phase. Additionally, the level of performance achieved with the same number of stages as a result of additional orbital slots, as reflected between Models U1 and U2, may allow for a lower transfer cost budget to be implemented for similar results, as long as an adequate number of orbital slots are provided.

A tertiary observation drawn from the results is the trends that often lead to Model~A being the highest-performing model in some cases. As shown in Table~\ref{tab:cross_reference_performance}, Model~A outperforms Model~U1 in 27 TCs and outperforms Model~U2 in five TCs; these TCs are shown in Appendix~C. A majority of the TCs in which Model~A was the highest-performing model have a long time duration, where 19 out of 27 (roughly \SI{70}{\%}) total TCs where Model~A outperformed Model~U1, and four out of five (\SI{80}{\%}) total TCs where Model~A outperformed Model~U2 have a time duration longer than the average time duration of all TCs, respectively. In general, this trend leads to high levels of performance from Model~A because as the time duration of the TC increases, the time between stages (for an equivalent number of stages) increases proportionally with the time duration, while the time step size between attitude control opportunities, $\Delta \tau$, remains constant. As such, this allows Model~A to perform more attitude control opportunities since the time step size of $\Delta \tau = \SI{1800}{s}$ is unchanged between differing TCs, while the reconfigurability models maintain the number of stages rather than the interval between them. Therefore, two TCs with different time durations will have proportionally different intervals between stages. This difference in the opportunities for control and the amount of time between them in Model~A versus the reconfigurability models allows for more fine control in Model~A, thus potentially performing better. However, the amount of TC characteristics that may additionally contribute to agility performance is extensive and should be investigated further in future work.  

Various parameter selections may be improved for a more thorough comparison of the CONOPS modeled in this paper. Firstly, the rate at which slewing opportunities occur in the agility model, $\Delta \tau = \SI{1800}{seconds}$ may be increased to allow more frequent attitude control and possibly more optimal pointing directions. Secondly, the constellation reconfigurability models assume equally spaced stages throughout the mission duration, while variably spaced stages may result in more optimal constellation reconfigurability. Finally, the cost computation algorithms used to compute the cost matrix assume high-thrust impulse maneuvers, while other transfer algorithms and trajectory optimization algorithms are present in literature such as Refs.~\cite{Hughes2003,Luo2007}, in addition to leveraging the secular changes of RAAN due to natural $J_2$ perturbations, could be employed to ensure the cost of constellation reconfiguration is economically feasible. 

In general, these findings and the data presented contribute to the accomplishment of the objectives previously established. Firstly, the comparison of all CONOPS with respect to the TCs selected provides a benchmark of performance for satellite agility and constellation reconfigurability with respect to nadir-directional systems. Secondly, the wide range of parameter combinations utilized in the reconfigurability models provides a further investigation into the emergent CONOPS of constellation reconfigurability. Thirdly, the high level of performance provided by the reconfigurability models establishes that constellation reconfigurability is worthy of a place in EO for application to TCs, additionally providing possible extensions to monitoring other short-term, rapid, and dynamic events. It is important to note that the results and conclusions drawn from this comparative analysis are not to be considered definitive, and should not be extended to other scenarios or applications. The main contribution of these results is the important trends regarding the performance of each CONOPS concerning a data set of varied historical TC characteristics.

\section{Conclusions} \label{sec:conclusion}

This paper sought to compare the CONOPS of satellite agility and constellation reconfigurability in order to provide a benchmark of each against a nadir-directional baseline and to determine the value of constellation reconfigurability in application to TC monitoring. The figure of merit---the number of quality observations---is obtained for each CONOPS with respect to 100 historical TCs. 

Overall, the comparison of each CONOPS performance as shown in Sec.~\ref{subsec:results} demonstrates that constellation reconfigurability has the capability to outperform satellite agility in application to TCs if provided an adequate number of stages for reconfiguration and high-quality destination orbital slots. Moreover, this capability applies both to {constellation reconfigurability} restricted to phasing maneuvers and one allowing plane change maneuvers. The increase in performance is especially emphasized through the application of the rapid nature of TCs, where these characteristics are paramount to optimal constellation configurations. Additionally, it is important to note that satellite agility may outperform constellation reconfigurability in cases where the additional fine control is beneficial, which trends toward longer TCs. However, there are extensive TC characteristics that may influence the performance of satellite agility versus constellation reconfigurability which are worthy of future research. As such, a direct conclusion regarding which CONOPS is more effective in monitoring TCs cannot be explicitly stated, but the provided trend in the performance of constellation reconfigurability lends validity that such a CONOPS may prove useful in EO applications.

There are many fruitful avenues for future research in the comparison between various CONOPS. Firstly, application to an expanded set of TCs either in the same manner as this paper or with multiple TCs in series or parallel may provide further insights into the trends presented in this paper. In conjunction, future research may apply additional natural disaster types such as flooding, earthquakes, volcanic eruptions, wildfires, or tsunamis to rigorously compare the concepts of satellite agility and constellation reconfigurability. Furthermore, the investigation of additional sensor types with a variety of FOV values may prove useful in the evaluation of additional data types. Additionally, future work may incorporate trajectory optimization algorithms to lower the fuel cost of reconfiguration and maintain economic feasibility, as mentioned in Sec.~\ref{subsec:results}. In conjunction with trajectory optimization, a study relating to the trade-off between performance and cost to determine the condition in which satellite agility proves to be more economically viable than constellation reconfigurability would be most insightful. Finally, the addition of constellation reconfigurability to the Earth observation satellite scheduling problem (EOSSP) in a similar manner to the agile EOSSP may prove insightful.

\section*{Appendix A: Rotation Matrix Computation}

The agility model depicted in Sec.~\ref{subsec:agile} requires a rotation matrix, $M$, to compute the current pointing direction $\bm{D}$ from the nadir direction $\bm{N}$ and Euler angles $\alpha,~\beta$, and $\gamma$. $M$ is composed of three individual matrices aligned with rotation about a different axis, in this case, ordered $\hat{x},~\hat{y}$, and $\hat{z}$ as $M_{x}(\alpha),~M_{y}(\beta)$, and $M_{z}(\gamma)$, calculated as:

\begin{equation}
    M_{x}(\alpha) = \begin{bmatrix}
    1 & 0 & 0 \\
    0 & \cos(\alpha) & \sin(\alpha)\\
    0 & -\sin(\alpha) & \cos(\alpha)
                    \end{bmatrix}, \quad
    M_{y}(\beta) = \begin{bmatrix}
    \cos(\beta) & 0 & -\sin(\beta)\\
    0 & 1 & 0 \\
    \sin(\beta) & 0 & \cos(\beta)
                    \end{bmatrix}, \quad
    M_{z}(\gamma) = \begin{bmatrix}
    \cos(\gamma) & \sin(\gamma) & 0\\
    -\sin(\gamma) & \cos(\gamma) & 0 \\
    0 & 0 & 1
                    \end{bmatrix}
\label{eq:Rx,y,z}
\end{equation}

Finally, the matrices $M_{x}(\alpha),~M_{y}(\beta)$, and $M_{z}(\gamma)$ are multiplied in the order of rotation about $\hat{x}$, then about $\hat{y}$, then about $\hat{z}$ to compute $M$ as shown in Eq.~\eqref{eq:Rrot}. All rotation matrices are derived from those in Ref.~\cite{Lesk1986Rrot}.

\begin{equation}
    \begin{split}
    M &= M_{x}(\alpha) M_{y}(\beta) M_{z}(\gamma) = \\
    &\begin{bmatrix}
        \cos(\beta)\cos(\gamma) & \cos(\beta)\sin(\gamma) & -\sin(\beta) \\
        \sin(\alpha)\sin(\beta)\cos(\gamma) - \cos(\alpha)\sin(\alpha) & \sin(\alpha)\sin(\beta)\sin(\gamma) + \cos(\alpha)\cos(\gamma) & \cos(\alpha)\cos(\beta) \\
        \cos(\alpha)\sin(\beta)\cos(\gamma) + \sin(\alpha)\sin(\gamma) & \cos(\alpha)\sin(\beta)\sin(\gamma) - \sin(\alpha)\cos(\gamma) & -\sin(\alpha)\cos(\beta) 
    \end{bmatrix}
    \end{split}
    \label{eq:Rrot}
\end{equation}

\section*{Appendix B: 30-Degree FOV Case}
\begin{figure}[!ht]
    \centering
    \includegraphics[width = \textwidth]{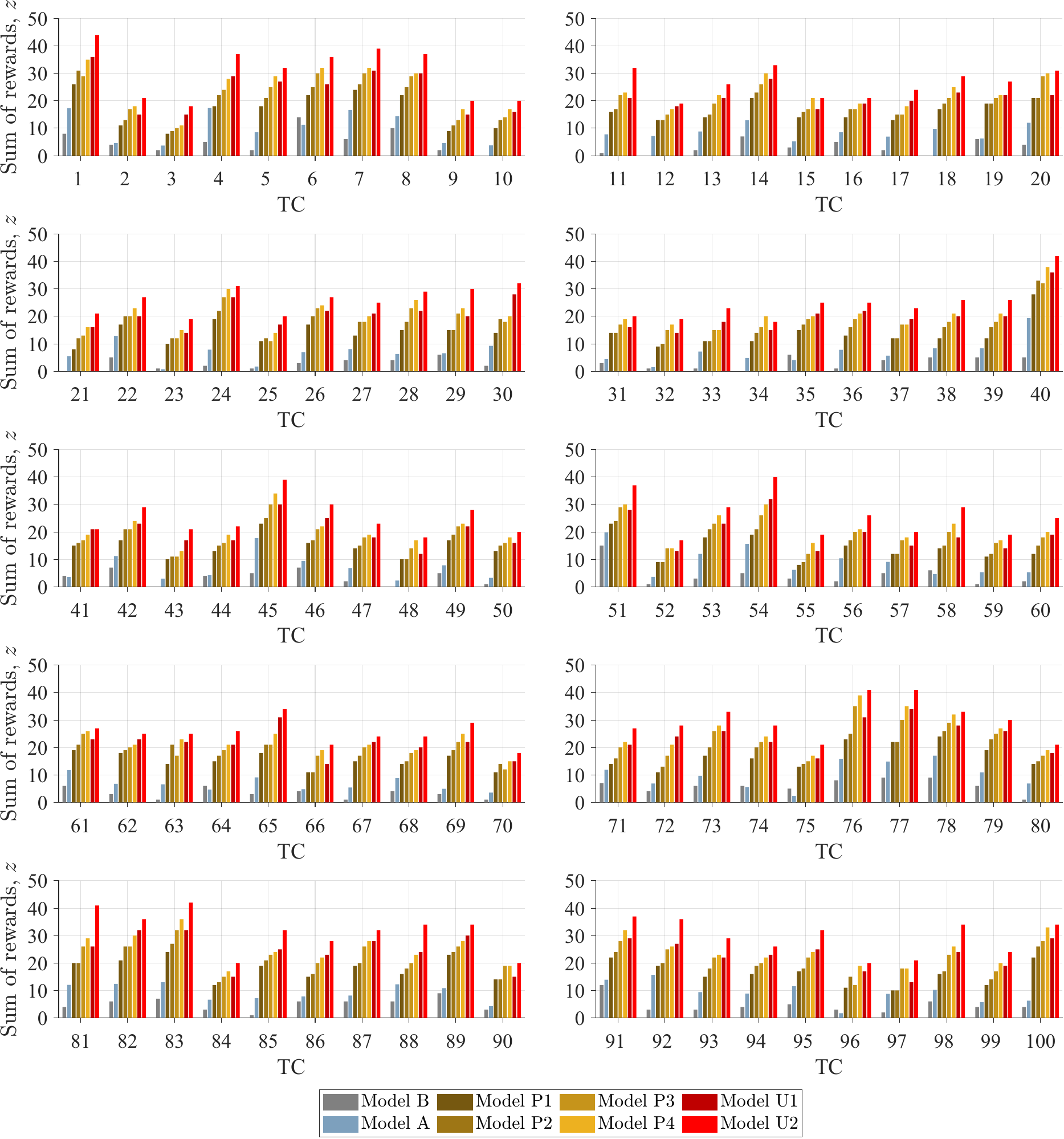}
    \caption{Sum of observation rewards, $z$, with respect to a FOV of 30 degrees for all TCs.}
    \label{fig:Tempest_rewards_NewFOV}
\end{figure}

\begin{figure}[!ht]
    \centering
    \includegraphics[width=0.6\linewidth]{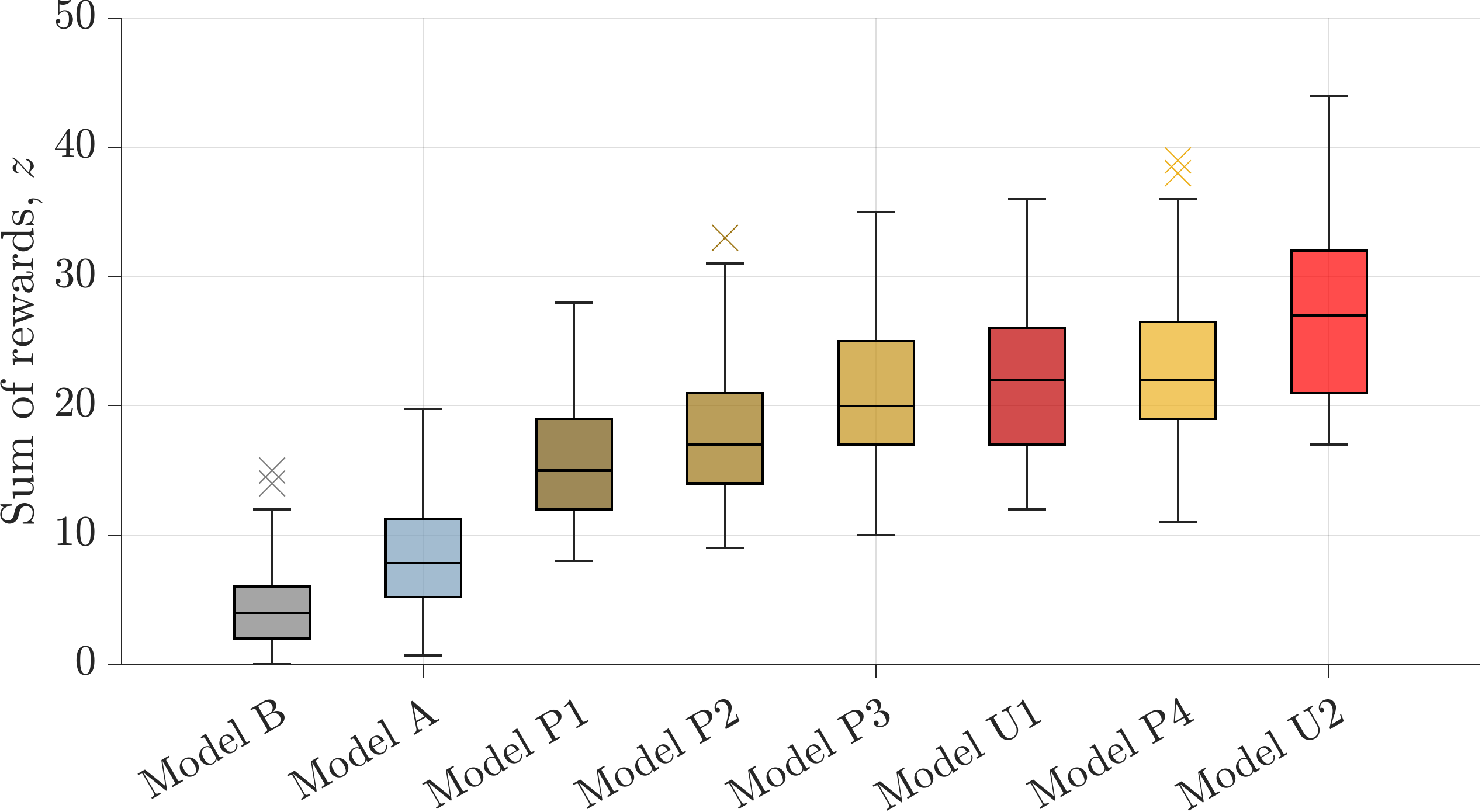}
    \caption{Sum of rewards organized by average performance with respect to a FOV of 30 degrees.}
    \label{fig:box_Z_NewFOV}
\end{figure}

The results of simulating each CONOPS model subject to identical parameters from the 45-degree FOV case presented in Sec.~\ref{subsec:results}, with a reduction in the FOV to 30 degrees, are provided in this appendix. Figure~\ref{fig:Tempest_rewards_NewFOV} depicts a summary of the sum of observation rewards, $z$, with an individual grouping for each TC. All figure and table formatting are identical to those in Sec.~\ref{subsec:results}. In comparison to the 45-degree FOV case, all CONOPS performed less as a result of the reduced FOV, in some cases driving Model~B to have zero observations. Additionally, Fig.~\ref{fig:box_Z_NewFOV} displays the sum of observation rewards, $z$, as a box chart. However, the ordering is slightly changed, wherein Model~A performs less than Model~P1 on average and Model~U1 performs less than Model~P4 on average. Similarly to the 45-degree FOV case, Model~U2 is the highest-performing model on average.

Furthermore, in the order that each model appears in Fig.~\ref{fig:box_Z_NewFOV}, Table~\ref{tab:Performance_over_baseline_NewFOV} provides statistics on the increase in performance over Model~B. Key points in the comparison of Tables~\ref{tab:Performance_over_baseline_NewFOV} and~\ref{tab:Performance_over_baseline} include the different ordering of each model, the increased change in performance relative to Model~B, and the negative improvement of Model~A. Another crucial point is that Model~U2 remains the highest-performing model on average with the highest minimum and maximum increase, while Model~A is now the lowest-performing model on average also with the lowest minimum and maximum increase.

\begin{table}[!ht]
    \centering
    \caption{Percent increase in performance of all models over Model~B with respect to a FOV of 30 degrees.}
    \begin{tabular}{ l r r r }
         \hline \hline 
         Model & Average $\pm$ standard deviation, \% & Minimum, \% & Maximum, \% \\
         \hline 
         Model~A  & $ 153.61 \pm 169.44 $ & -53.09 & 681.16 \\
         Model~P1 & $ 419.68 \pm 364.54 $ & 53.33  & 1800   \\
         Model~P2 & $ 491.32 \pm 427.51 $ & 60     & 2000   \\
         Model~P3 & $ 583.26 \pm 475.11 $ & 93.33  & 2200   \\
         Model~U1 & $ 637.96 \pm 525.92 $ & 85.71  & 2400   \\
         Model~P4 & $ 664.27 \pm 525.39 $ & 100    & 2300   \\
         Model~U2 & $ 825.53 \pm 650.84 $ & 146.67 & 3100   \\
         \hline \hline 
    \end{tabular}
    \label{tab:Performance_over_baseline_NewFOV}
\end{table}

In addition to Model~U2 being the highest-performing model over Model~B, Model~U2 outperforms all other models in nearly all (98 or more) TCs. Table~\ref{tab:cross_reference_performance_NewFOV} provides the number of TCs in which the model in each column outperforms the model in each row, reflecting that while Model~U2 is the highest-performing on average Model~P4 additionally outperforms other models in a majority of TCs. Specifically, Model~P4 outperforms all models other than Model~U2 in 64 or more TCs, as well as outperforming Model~U2 in one TC. Furthermore, Model~U1 outperforms all other models with the exception of Models~P4 and~U2 in 62 or more TCs. As such, the reconfigurability models with high numbers of stages or high flexibility of orbital slots are the highest-performing models. Additionally, Model~B outperforms Model~A in nine TCs, and Model~A does not outperform any reconfigurability models in any TCs. 

\begin{table}[!ht]
    \centering
    \caption{The number of TCs with respect to a FOV of 30 degrees in which the model of a column outperforms the model in the row.}
    \resizebox{\textwidth}{!}{
    \begin{tabular}{ l r r r r r r r r}
         \hline \hline
         & Model~B & Model~A & Model~P1 & Model~P2 & Model~P3 & Model~P4 & Model~U1 & Model~U2 \\
         \hline 
         Model~B  & - & 91 & 100 & 100 & 100 & 100 & 100 & 100 \\
         Model~A  & 9 & -  & 100 & 100 & 100 & 100 & 100 & 100 \\
         Model~P1 & 0 & 0  & -   & 85  & 99  & 100 & 100 & 100 \\
         Model~P2 & 0 & 0  & 0   & -   & 84  & 100 & 99  & 100 \\
         Model~P3 & 0 & 0  & 0   & 7   & -   & 95  & 62  & 100 \\
         Model~P4 & 0 & 0  & 0   & 0   & 0   & -   & 29  & 98  \\
         Model~U1 & 0 & 0  & 0   & 0   & 22  & 64  & -   & 99  \\
         Model~U2 & 0 & 0  & 0   & 0   & 0   & 1   & 0   & -   \\
         \hline \hline
    \end{tabular}
    }
    \begin{tablenotes}
        \item[-] A hyphen (-) indicates that the models are the same, and thus cannot outperform itself.
    \end{tablenotes}
    \label{tab:cross_reference_performance_NewFOV}
\end{table}

Through the data presented concerning a reduced FOV of 30 degrees, most observations from the 45-degree FOV case are further reinforced. Firstly, the observation that constellation reconfigurability has the potential to outperform satellite agility in response to TCs when given adequate parameters is reinforced through Models~U2 and P4 functioning as the two highest-performing models. In addition to Tables~\ref{tab:Performance_over_baseline_NewFOV} and~\ref{tab:cross_reference_performance_NewFOV}, Table~\ref{tab:ModelU2_performance_NewFOV} depicts the performance of Model~U2 above each other model, reflecting that Model~U2 at least achieves identical performance to all other models, with the exception of one case from Model~P4. The performance from these reconfigurability models again mainly results from frequent reconfiguration opportunities and the flexibility provided by the orbital slot options, as with Model~U2. Secondly, the observation that orbital slot flexibility is more important than the number of stages for reconfiguration is reinforced through Model~U2 again being the highest-performing model, outperforming the second highest-performing Model~P4 in 98 TCs while having the same number of stages but enabling plane change maneuvers as well as phasing maneuvers. However, the observation that Model~A performs better in some cases is not reinforced by the reduced FOV cases. 

\begin{table}[!ht]
    \centering
    \caption{Percent increase in performance of Model~U2 over other models with respect to a FOV of 30 degrees.}
    \begin{tabular}{ l r r r }
         \hline \hline 
         Model & Average $\pm$ standard deviation, \% & Minimum, \% & Maximum, \% \\
         \hline 
         Model~A  & $ 319.44 \pm 312.85 $ & 87.07 & 2750.50 \\
         Model~P1 & $ 79.78  \pm 25.44  $ & 37.50 & 162.50  \\
         Model~P2 & $ 59.04  \pm 22.95  $ & 19.05 & 115.38  \\
         Model~P3 & $ 36.09  \pm 16.76  $ & 5.26  & 90.91   \\
         Model~P4 & $ 20.83  \pm 13.02  $ & -10   & 63.64   \\
         Model~U1 & $ 26.63  \pm 11.51  $ & 0     & 61.54   \\
         \hline \hline 
    \end{tabular}
    \label{tab:ModelU2_performance_NewFOV}
\end{table}

One major difference in the results is the highly reduced performance of Model~A, as reflected in Table~\ref{tab:cross_reference_performance_NewFOV}. One aspect that may lead to this decrease in performance is the reduction in FOV, leading to a smaller number of visibility occurrences in general. Additionally, since the FOV is smaller, each satellite may be required to slew further to view a target, which directly penalizes the quality of the rewards gained according to Eq.~\ref{eq:Degredation}. Therefore, even in the event that slewing would provide a relatively high number of visibility occurrences, the quality penalty caused by the intense slewing angles impacts the observation rewards enough to perform worse than the other models. In combination, these two aspects reduce the number and quality of rewards gained by Model~A, thus causing Model~A to perform significantly worse than when provided a larger FOV of 45 degrees. 

In general, these additional findings further solidify previous findings. Firstly, the reduced FOV case provides an additional benchmark of performance for satellite agility and constellation reconfigurability with respect to yet another changed parameter. Secondly, the level of performance maintained by the reconfigurability models despite the reduction in the FOV reinforces that constellation reconfigurability is worthy of a place in EO for application to TCs. Again, it is important to note that the results and conclusions drawn from this additional FOV case are not to be considered definitive, and should not be extended to other applications. 

\section*{Appendix C: All TCs Where Model~A Outperforms Models~U1 and U2}

\begin{figure}[!ht]
    \centering
    \begin{subfigure}[h]{0.49\textwidth}
        \centering
        \includegraphics[width = \textwidth]{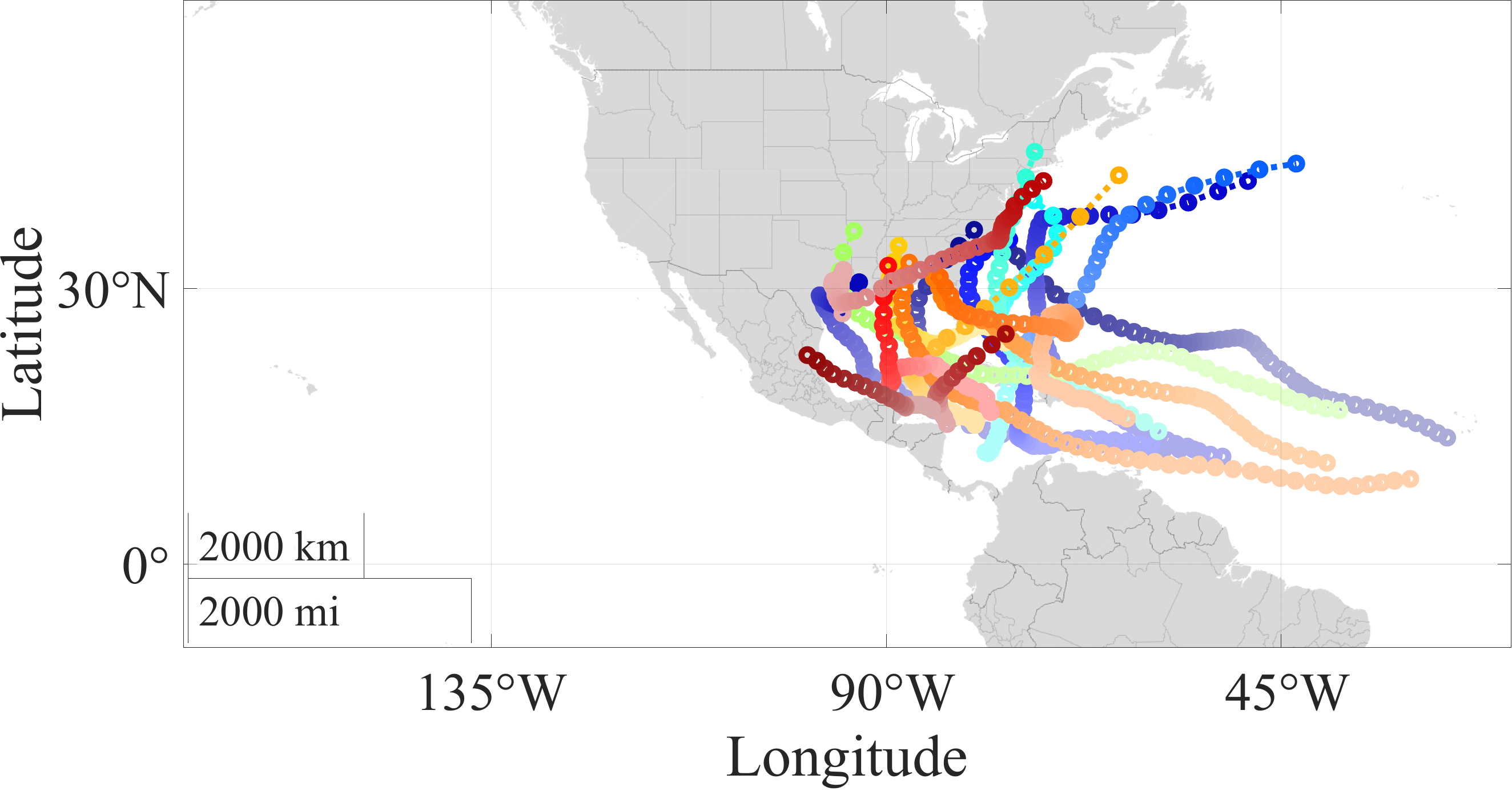}
        \caption{Hurricanes where Model~A outperforms Model~U1.}
        \label{fig:Hurricane_out_U1}
    \end{subfigure}
    \hfill
    \begin{subfigure}[h]{0.49\textwidth}
        \centering
        \includegraphics[width = \textwidth]{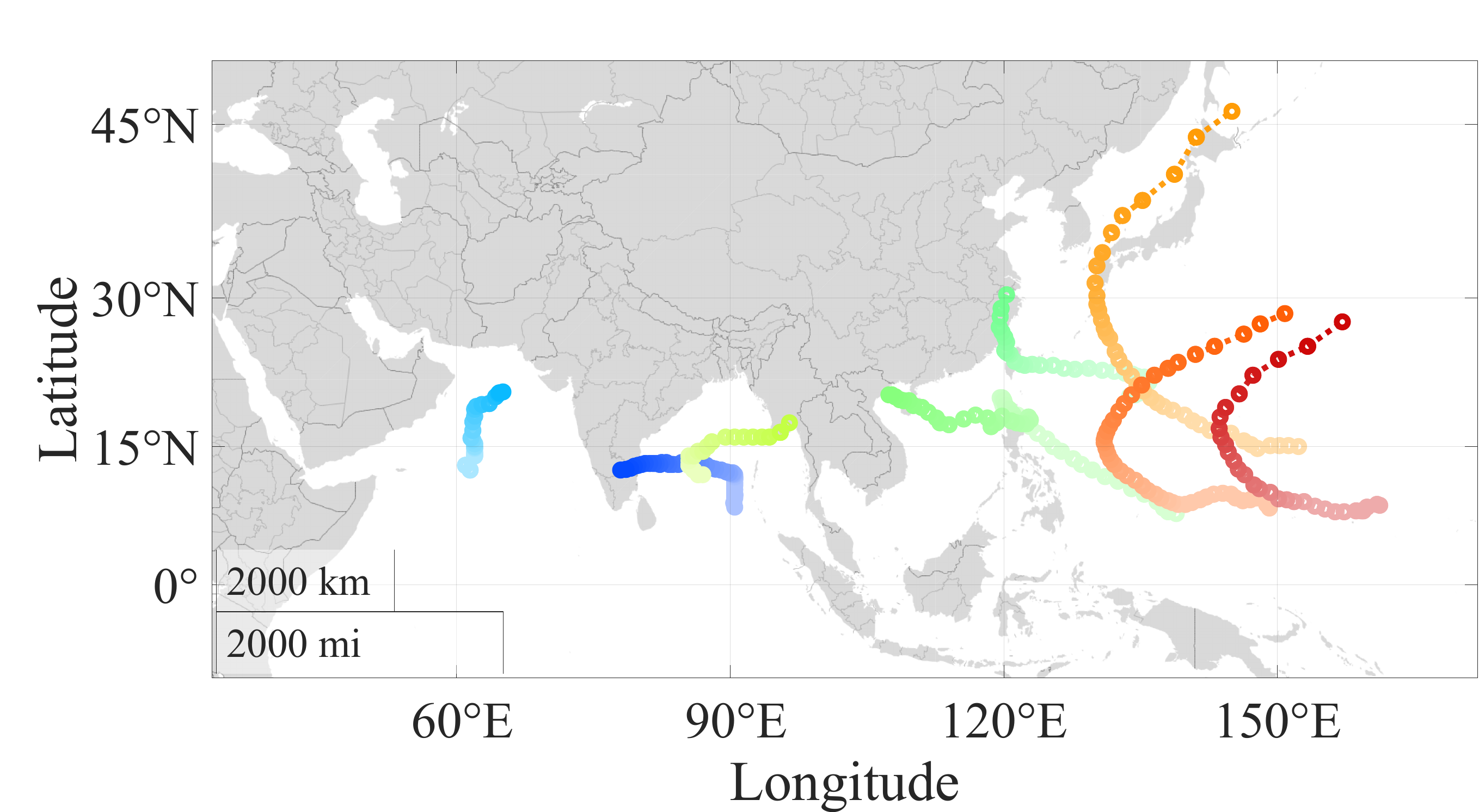}
        \caption{Typhoons where Model~A outperforms Model~U1.}
        \label{fig:Typhoon_out_U1}
    \end{subfigure}
    \begin{subfigure}[h]{0.49\textwidth}
        \centering
        \includegraphics[width = \textwidth]{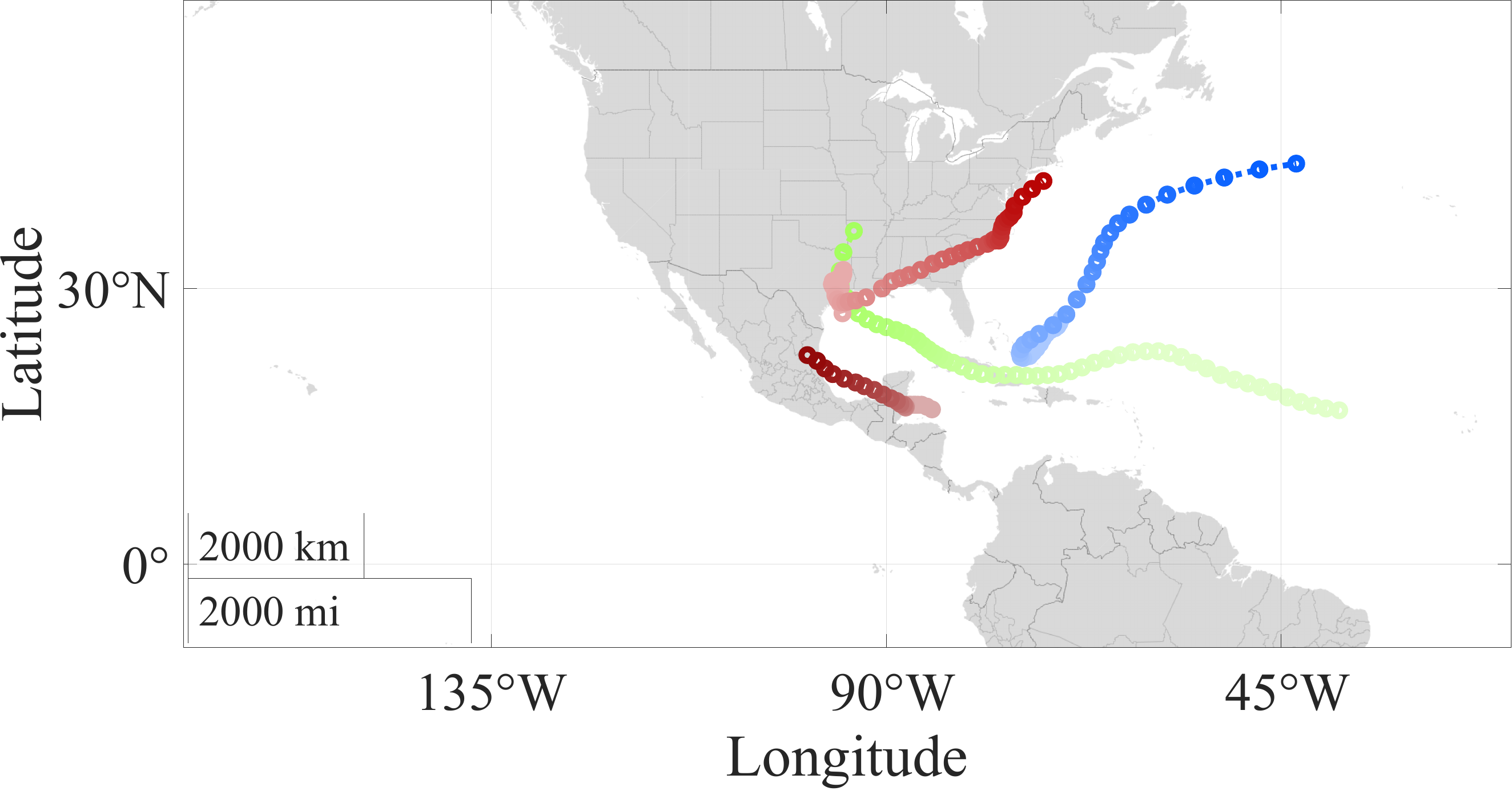}
        \caption{Hurricanes where Model~A outperforms Model~U2.}
        \label{fig:Hurricane_out_U2}
    \end{subfigure}
    \hfill
    \begin{subfigure}[h]{0.49\textwidth}
        \centering
        \includegraphics[width = \textwidth]{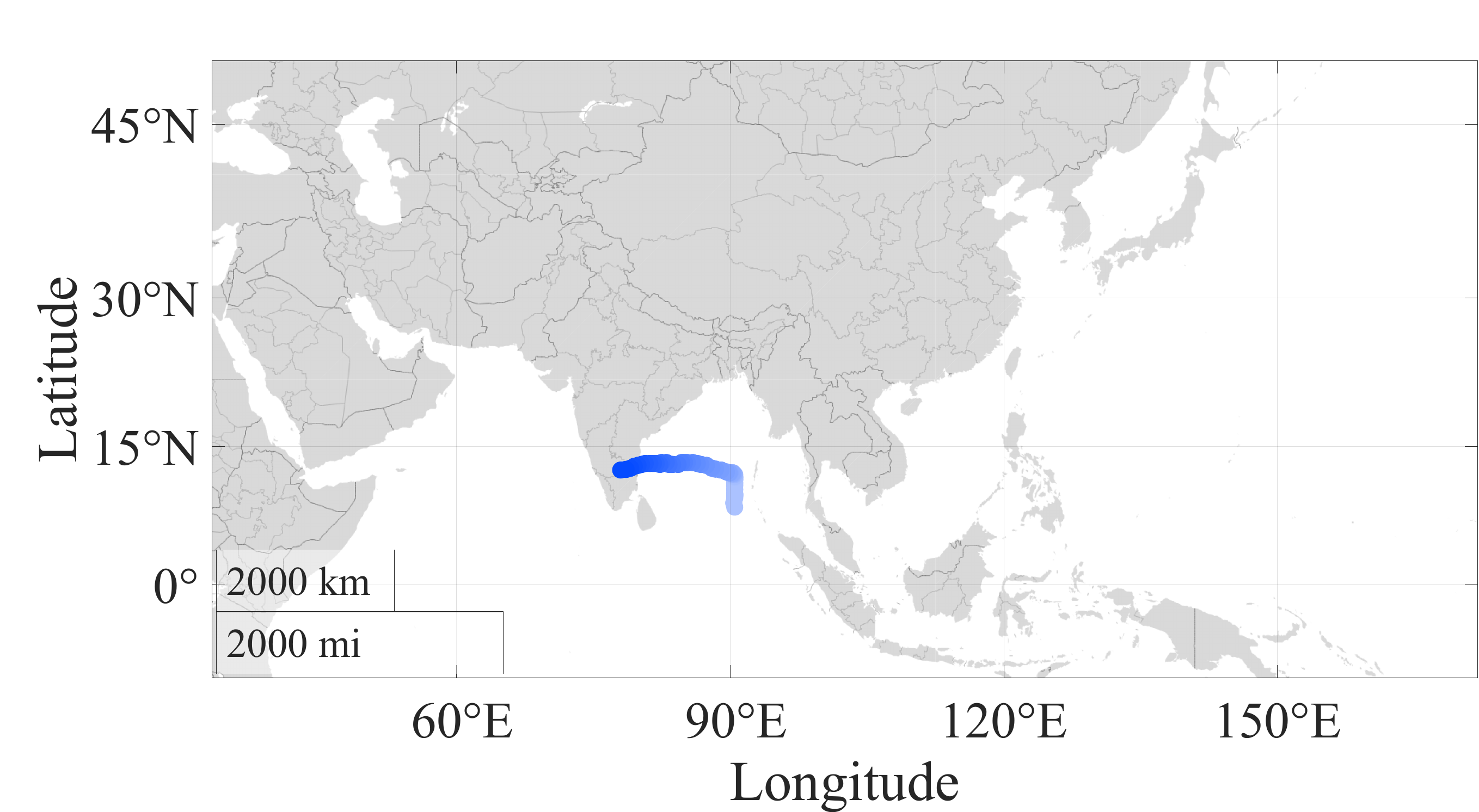}
        \caption{Typhoons where Model~A outperforms Model~U2.}
        \label{fig:Typhoon_out_U2}
    \end{subfigure}
    \caption{TCs where Model~A outperforms the highest performing reconfigurability models.}
    \label{fig:TC_out_agile}
\end{figure}

The TCs in which Model~A outperforms Models~U1 and~U2 in the 45-degree FOV case presented in Sec.~III.C are shown in Fig.~\ref{fig:TC_out_agile}. Figures~\ref{fig:Hurricane_out_U1} and~\ref{fig:Typhoon_out_U1} depicts the 19 Hurricanes and eight Typhoons where Model~A outperforms Model~U1, respectively. Additionally, Figs.~\ref{fig:Hurricane_out_U2} and~\ref{fig:Typhoon_out_U2} depicts the four Hurricanes and one Typhoon where Model~A outperforms Model~U2, respectively.

\clearpage
\newpage
\bibliography{references}

\begin{thebibliography}{45}
\newcommand{\enquote}[1]{``#1''}
\providecommand{\natexlab}[1]{#1}
\providecommand{\url}[1]{\texttt{#1}}
\providecommand{\urlprefix}{URL }
\expandafter\ifx\csname urlstyle\endcsname\relax
  \providecommand{\doi}[1]{\discretionary{}{}{}https://doi.org/#1}\else
  \providecommand{\doi}[1]{\discretionary{}{}{}\urlstyle{rm}\url{https://doi.org/#1}}\fi

\bibitem[{Gierach and Subrahmanyam(2007)}]{Gierach2007}
Gierach, M.~M., and Subrahmanyam, B., \enquote{Satellite Data Analysis of the Upper Ocean Response to Hurricanes Katrina and Rita (2005) in the Gulf of Mexico,} \emph{IEEE Geoscience and Remote Sensing Letters}, Vol.~4, No.~1, 2007, pp. 132--136.
\newblock \doi{10.1109/LGRS.2006.887145}.

\bibitem[{Pérez-Alarcón et~al.(2021)Pérez-Alarcón, Sorí, Fernández-Alvarez, Nieto, and Gimeno}]{Perez2021size}
Pérez-Alarcón, A., Sorí, R., Fernández-Alvarez, J.~C., Nieto, R., and Gimeno, L., \enquote{Comparative climatology of outer tropical cyclone size using radial wind profiles,} \emph{Weather and Climate Extremes}, Vol.~33, 2021, p. 100366.
\newblock \doi{10.1016/j.wace.2021.100366}.

\bibitem[{Brunkard et~al.(2008)Brunkard, Namulanda, and Ratard}]{Brunkard2008}
Brunkard, J., Namulanda, G., and Ratard, R., \enquote{Hurricane Katrina Deaths, Louisiana, 2005,} \emph{Disaster Medicine and Public Health Preparedness}, Vol.~2, No.~4, 2008, pp. 215--223.
\newblock \doi{10.1097/DMP.0b013e31818aaf55}.

\bibitem[{Burton and Hicks(2005)}]{Burton2005}
Burton, M., and Hicks, M., \enquote{Hurricane Katrina: Preliminary Estimates of Commercial and Public Sector Damages,} \emph{Center for Business and Economic Research, Marshall University}, 2005.

\bibitem[{Casey-Lockyer and Heick(2013)}]{Lockyer2018}
Casey-Lockyer, M., and Heick, R.~J., \enquote{Deaths Associated with Hurricane Sandy — October–November 2012,} \emph{Disaster Medicine and Public Health Preparedness}, Vol.~62, No.~20, 2013, pp. 393--397.

\bibitem[{Diakakis et~al.(2015)Diakakis, Deligiannakis, Katsetsiadou, and Lekkas}]{Diakakis2015}
Diakakis, M., Deligiannakis, G., Katsetsiadou, K., and Lekkas, E., \enquote{Hurricane Sandy mortality in the Caribbean and continental North America,} \emph{Disaster Prevention and Management}, Vol.~24, No.~1, 2015, pp. 132--148.
\newblock \doi{10.1108/DPM-05-2014-0082}.

\bibitem[{Kunz et~al.(2013)Kunz, Mühr, Kunz-Plapp, Daniell, Khazai, Wenzel, Vannieuwenhuyse, Comes, Elmer, Schröter, and et~al.}]{Kunz2013}
Kunz, M., Mühr, B., Kunz-Plapp, T., Daniell, J.~E., Khazai, B., Wenzel, F., Vannieuwenhuyse, M., Comes, T., Elmer, F., Schröter, K., and et~al., \enquote{Investigation of superstorm sandy 2012 in a multi-disciplinary approach,} \emph{Natural Hazards and Earth System Sciences}, Vol.~13, No.~10, 2013, p. 2579–2598.
\newblock \doi{10.5194/nhess-13-2579-2013}.

\bibitem[{Singaravelu(2013)}]{Raghavan2013}
Singaravelu, R., \enquote{Observational aspects including weather radar for tropical cyclone monitoring,} \emph{Mausam}, Vol.~64, 2013, pp. 89--96.
\newblock \doi{10.54302/mausam.v64i1.658}.

\bibitem[{Holbach et~al.(2023)Holbach, Bousquet, Bucci, Chang, Cione, Ditchek, Doyle, Duvel, Elston, Goni, Hon, Ito, Jelenak, Lei, Lumpkin, McMahon, Reason, Sanabia, Shay, Sippel, Sushko, Tang, Tsuboki, Yamada, Zawislak, and Zhang}]{HOLBACH2023}
Holbach, H.~M., Bousquet, O., Bucci, L., Chang, P., Cione, J., Ditchek, S., Doyle, J., Duvel, J.-P., Elston, J., Goni, G., Hon, K.~K., Ito, K., Jelenak, Z., Lei, X., Lumpkin, R., McMahon, C.~R., Reason, C., Sanabia, E., Shay, L.~K., Sippel, J.~A., Sushko, A., Tang, J., Tsuboki, K., Yamada, H., Zawislak, J., and Zhang, J.~A., \enquote{Recent advancements in aircraft and in situ observations of tropical cyclones,} \emph{Tropical Cyclone Research and Review}, Vol.~12, No.~2, 2023, pp. 81--99.
\newblock \doi{10.1016/j.tcrr.2023.06.001}.

\bibitem[{Amarin et~al.(2012)Amarin, Jones, El-Nimri, Johnson, Ruf, Miller, and Uhlhorn}]{Amarin2012}
Amarin, R.~A., Jones, W.~L., El-Nimri, S.~F., Johnson, J.~W., Ruf, C.~S., Miller, T.~L., and Uhlhorn, E., \enquote{Hurricane wind speed measurements in rainy conditions using the Airborne Hurricane Imaging Radiometer (Hirad),} \emph{IEEE Transactions on Geoscience and Remote Sensing}, Vol.~50, No.~1, 2012, p. 180–192.
\newblock \doi{10.1109/tgrs.2011.2161637}.

\bibitem[{Brown et~al.(2017)Brown, Focardi, Kitiyakara, Maiwald, Milligan, Montes, Padmanabhan, Redick, Russel, Bach, and Walkemeyer}]{COWVR-7943884}
Brown, S., Focardi, P., Kitiyakara, A., Maiwald, F., Milligan, L., Montes, O., Padmanabhan, S., Redick, R., Russel, D., Bach, V., and Walkemeyer, P., \enquote{The COWVR Mission: Demonstrating the capability of a new generation of small satellite weather sensors,} \emph{2017 IEEE Aerospace Conference}, 2017, pp. 1--7.
\newblock \doi{10.1109/AERO.2017.7943884}.

\bibitem[{Reising et~al.(2018)Reising, Gaier, Padmanabhan, Lim, Heneghan, Kummerow, Berg, Chandrasekar, Radhakrishnan, Brown, Carvo, and Pallas}]{TEMPEST-8517330}
Reising, S.~C., Gaier, T.~C., Padmanabhan, S., Lim, B.~H., Heneghan, C., Kummerow, C.~D., Berg, W., Chandrasekar, V., Radhakrishnan, C., Brown, S.~T., Carvo, J., and Pallas, M., \enquote{An Earth Venture In-Space Technology Demonstration Mission for Temporal Experiment for Storms and Tropical Systems (Tempest),} \emph{IGARSS 2018 - 2018 IEEE International Geoscience and Remote Sensing Symposium}, 2018, pp. 6301--6303.
\newblock \doi{10.1109/IGARSS.2018.8517330}.

\bibitem[{Wilson et~al.(2018)Wilson, Angal, and Xiong}]{Wilson2018MODIS}
Wilson, T.~M., Angal, A., and Xiong, X., \enquote{Sensor Performance Assessment for Terra and Aqua Modis using unscheduled lunar observations,} \emph{Sensors, Systems, and Next-Generation Satellites XXII}, 2018.
\newblock \doi{10.1117/12.2324873}.

\bibitem[{Rose et~al.(2013)Rose, Ruf, Rose, Brummitt, and Ridley}]{Rose2013}
Rose, R., Ruf, C., Rose, D., Brummitt, M., and Ridley, A., \enquote{The CYGNSS Flight Segment; a major NASA science mission enabled by micro-satellite Technology,} \emph{2013 IEEE Aerospace Conference}, 2013.
\newblock \doi{10.1109/aero.2013.6497205}.

\bibitem[{Boustan et~al.(2020)Boustan, Kahn, Rhode, and Yanguas}]{boustan2020}
Boustan, L.~P., Kahn, M.~E., Rhode, P.~W., and Yanguas, M.~L., \enquote{The effect of natural disasters on economic activity in US counties: A century of data,} \emph{Journal of Urban Economics}, Vol. 118, 2020, p. 103257.
\newblock \doi{10.1016/j.jue.2020.103257}.

\bibitem[{Haiquan et~al.(2019)Haiquan, Wei, Xiaoxuan, and Chongyan}]{SUN2019EOSSET}
Haiquan, S., Wei, X., Xiaoxuan, H., and Chongyan, X., \enquote{Earth observation satellite scheduling for emergency tasks,} \emph{Journal of Systems Engineering and Electronics}, Vol.~30, No.~5, 2019, pp. 931--945.
\newblock \doi{10.21629/JSEE.2019.05.11}.

\bibitem[{Chatterjee and Tharmarasa(2022)}]{CHATTERJEE2022}
Chatterjee, A., and Tharmarasa, R., \enquote{Reward Factor-Based Multiple Agile Satellites Scheduling With Energy and Memory Constraints,} \emph{IEEE Transactions on Aerospace and Electronic Systems}, Vol.~58, No.~4, 2022, pp. 3090--3103.
\newblock \doi{10.1109/TAES.2022.3146115}.

\bibitem[{Chen et~al.(2020)Chen, Xu, Shen, Zhang, Zezhong, and Xu}]{Chen2020}
Chen, Y., Xu, M., Shen, X., Zhang, G., Zezhong, L., and Xu, J., \enquote{A Multi-Objective Modeling Method of Multi-Satellite Imaging Task Planning for Large Regional Mapping,} \emph{Remote Sensing}, Vol.~12, 2020, p. 344.
\newblock \doi{10.3390/rs12030344}.

\bibitem[{Lemaitre et~al.(2002)Lemaitre, Verfaillie, Jouhaud, Lachiver, and Bataille}]{LEMAITRE2002}
Lemaitre, M., Verfaillie, G., Jouhaud, F., Lachiver, J.-M., and Bataille, N., \enquote{Selecting and scheduling observations of agile satellites,} \emph{Aerospace Science and Technology}, Vol.~6, No.~5, 2002, pp. 367--381.
\newblock \doi{10.1016/S1270-9638(02)01173-2}.

\bibitem[{Song et~al.(2018)Song, Chen, Luo, Wang, and Dai}]{Song2018}
Song, Z., Chen, X., Luo, X., Wang, M., and Dai, G., \enquote{Multi-objective optimization of Agile Satellite Orbit Design,} \emph{Advances in Space Research}, Vol.~62, No.~11, 2018, p. 3053–3064.
\newblock \doi{10.1016/j.asr.2018.08.037}.

\bibitem[{Peng et~al.(2020)Peng, Song, Xing, Gunawan, and Vansteenwegen}]{Peng2020timedependent}
Peng, G., Song, G., Xing, L., Gunawan, A., and Vansteenwegen, P., \enquote{An exact algorithm for Agile Earth observation satellite scheduling with time-dependent profits,} \emph{Computers and Operations Research}, Vol. 120, 2020, p. 104946.
\newblock \doi{10.1016/j.cor.2020.104946}.

\bibitem[{Lee et~al.(2024)Lee, Williams~Rogers, Pearl, Chen, and Ho}]{lee2024deterministic}
Lee, H., Williams~Rogers, D.~O., Pearl, B.~D., Chen, H., and Ho, K., \enquote{Deterministic Multistage Constellation Reconfiguration Using Integer Programming and Sequential Decision-Making Methods,} \emph{Journal of Spacecraft and Rockets}, Vol.~0, No.~0, 2024, pp. 1--17.
\newblock \doi{10.2514/1.A35990}.

\bibitem[{Morgan et~al.(2023)Morgan, McGrath, and de~Weck}]{Morgan2023}
Morgan, S.~J., McGrath, C.~N., and de~Weck, O.~L., \enquote{Optimization of multispacecraft maneuvers for mobile target tracking from low Earth orbit,} \emph{Journal of Spacecraft and Rockets}, Vol.~60, No.~2, 2023, p. 581–590.
\newblock \doi{10.2514/1.a35457}.

\bibitem[{Lee et~al.(2022)Lee, Chen, and Ho}]{Lee2022}
Lee, H., Chen, H., and Ho, K., \enquote{Maximizing Observation Throughput via Multi-Stage Satellite Constellation Reconfiguration,} \emph{AAS/AIAA Astrodynamics Specialist Conference}, 2022.

\bibitem[{Lee and Liu(2023)}]{Lee2023}
Lee, H., and Liu, Z., \enquote{A Novel Formulation for the Multi-Stage Satellite Constellation Reconfiguration Problem: Initial Results,} \emph{33rd AAS/AIAA Space Flight Mechanics Meeting}, 2023.

\bibitem[{Lee and Ho(2020)}]{Lee2020binary}
Lee, H., and Ho, K., \enquote{Binary Integer Linear Programming Formulation for Optimal Satellite Constellation Reconfiguration,} \emph{AAS/AIAA Astrodynamics Specialist Conference}, 2020.

\bibitem[{Lee and Ho(2021)}]{Lee2021lagrangian}
Lee, H., and Ho, K., \enquote{A Lagrangian Relaxation-Based Heuristic Approach to Regional Constellation Reconfiguration Problem,} \emph{AAS/AIAA Astrodynamics Specialist Conference}, 2021.

\bibitem[{Lee and Ho(2023)}]{lee2023regional}
Lee, H., and Ho, K., \enquote{Regional Constellation Reconfiguration Problem: Integer Linear Programming Formulation and Lagrangian Heuristic Method,} \emph{Journal of Spacecraft and Rockets}, Vol.~60, No.~6, 2023, pp. 1828--1845.
\newblock \doi{10.2514/1.A35685}.

\bibitem[{Branch et~al.(2023)Branch, Marchetti, Mason, Montgomery, Johnson, Chien, Wu, Smith, Mandrake, and Tavallali}]{Branch2023}
Branch, A., Marchetti, Y., Mason, J., Montgomery, J., Johnson, M.~C., Chien, S., Wu, L., Smith, B., Mandrake, L., and Tavallali, P., \enquote{Federating Planning of Observations for Earth Science,} \emph{Proc. of International Workshop on Planning and Scheduling for Space}, 2023.
\newblock \urlprefix\url{https://ai.jpl.nasa.gov/public/documents/papers/Branch-IWPSS2023-federated.pdf}.

\bibitem[{Nag et~al.(2020)Nag, Moghaddam, Selva, Frank, Ravindra, Levinson, Azemati, Aguilar, Li, and Akbar}]{Nag2020}
Nag, S., Moghaddam, M., Selva, D., Frank, J., Ravindra, V., Levinson, R., Azemati, A., Aguilar, A., Li, A., and Akbar, R., \enquote{D-shield: Distributed spacecraft with heuristic intelligence to enable logistical decisions,} \emph{IGARSS 2020 - 2020 IEEE International Geoscience and Remote Sensing Symposium}, 2020.
\newblock \doi{10.1109/igarss39084.2020.9323248}.

\bibitem[{Wang et~al.(2021)Wang, Wu, Xing, and Pedrycz}]{Wang2021}
Wang, X., Wu, G., Xing, L., and Pedrycz, W., \enquote{Agile Earth Observation Satellite Scheduling Over 20 Years: Formulations, Methods, and Future Directions,} \emph{IEEE Systems Journal}, Vol.~15, No.~3, 2021, pp. 3881--3892.
\newblock \doi{10.1109/JSYST.2020.2997050}.

\bibitem[{Pearl et~al.(2023)Pearl, Gold, and Lee}]{Pearl2023}
Pearl, B., Gold, L., and Lee, H., \enquote{Comparing the Effectiveness of Agility and Reconfigurability in Earth Observation Satellite Systems for Disaster Response,} \emph{AAS/AIAA Astrodynamics Specialist Conference}, 2023.

\bibitem[{Vallado(2013)}]{Vallado2013}
Vallado, D., \emph{Fundamentals of Astrodynamics and Applications}, Space technology library, Microcosm Press, 2013, Chap.~6, pp. 317--421.

\bibitem[{{National Hurricane Center} and {Central Pacific Hurricane Center}(2024)}]{RetiredNames}
{National Hurricane Center}, and {Central Pacific Hurricane Center}, \enquote{Tropical Cyclone Naming History and Retired Names,} , 2024.
\newblock \urlprefix\url{https://www.nhc.noaa.gov/aboutnames_history.shtml}, last Accessed September 29, 2024.

\bibitem[{{Weather Underground}(2024)}]{WUnderground}
{Weather Underground}, \enquote{Hurricane Archive,} , 2024.
\newblock \urlprefix\url{https://www.wunderground.com/hurricane/archive}, last Accessed September 29, 2024.

\bibitem[{US~Department~of Commerce(2022)}]{TempestDefinitions_2022}
US~Department~of Commerce, N., \enquote{Tropical definitions,} , May 2022.
\newblock \urlprefix\url{https://www.weather.gov/mob/tropical_definitions}, last Accessed September 29, 2024.

\bibitem[{Stephens et~al.(2003)Stephens, Cooksley, Da~Silva~Curiel, Boland, Jason, Northham, Brewer, Anzalchi, Newell, Underwood, Machin, Sun, and Sweeting}]{Stephens2003DMC}
Stephens, P., Cooksley, J., Da~Silva~Curiel, A., Boland, L., Jason, S., Northham, J., Brewer, A., Anzalchi, J., Newell, H., Underwood, C., Machin, S., Sun, W., and Sweeting, S., \enquote{Launch of the international Disaster Monitoring Constellation; the development of a novel international partnership in space,} \emph{International Conference on Recent Advances in Space Technologies (RAST)}, 2003, pp. 525 -- 535.
\newblock \doi{10.1109/RAST.2003.1303972}.

\bibitem[{N2YO.com(2023)}]{DMC_TLE}
N2YO.com, \enquote{Disaster Monitoring Satellites,} , 2023.
\newblock \urlprefix\url{https://www.n2yo.com/satellites/?c=8}, last Accessed September 29, 2024.

\bibitem[{MATLAB(2023)}]{MATLAB}
MATLAB, \emph{23.2.0.2459199 (R2023b) Update 5}, The MathWorks Inc., Natick, Massachusetts, 2023.

\bibitem[{L{\"{o}}fberg(2004)}]{YALMIP}
L{\"{o}}fberg, J., \enquote{YALMIP : A Toolbox for Modeling and Optimization in MATLAB,} \emph{In Proceedings of the CACSD Conference}, Taipei, Taiwan, 2004, pp. 284 -- 289.

\bibitem[{Lappas et~al.(2002)Lappas, Steyn, and Underwood}]{Lappas2002}
Lappas, V., Steyn, W., and Underwood, C., \enquote{Attitude control for small satellites using control moment gyros,} \emph{Acta Astronautica}, Vol.~51, 2002, pp. 101--111.
\newblock \doi{10.1016/S0094-5765(02)00089-9}.

\bibitem[{Karpenko and King(2019)}]{Karpenko2019}
Karpenko, M., and King, J.~T., \enquote{Maximizing agility envelopes for reaction wheel spacecraft,} \emph{Proceedings of the Institution of Mechanical Engineers, Part G: Journal of Aerospace Engineering}, Vol. 233, No.~8, 2019, pp. 2745--2759.
\newblock \doi{10.1177/0954410018787866}.

\bibitem[{Hughes et~al.(2003)Hughes, Mailhe, and Guzman}]{Hughes2003}
Hughes, S., Mailhe, L., and Guzman, J., \enquote{A Comparison of Trajectory Optimization Methods for the Impulsive Minimum Fuel Rendezvous Problem,} \emph{Advances in the Astronautical Sciences}, Vol. 113, 2003.

\bibitem[{Luo et~al.(2007)Luo, Lei, and Tang}]{Luo2007}
Luo, Y., Lei, Y., and Tang, G.-J., \enquote{Optimal Multi-Objective Nonlinear Impulsive Rendezvous,} \emph{Journal of Guidance, Control, and Dynamics}, Vol.~30, 2007, pp. 994--1002.
\newblock \doi{10.2514/1.27910}.

\bibitem[{Lesk(1986)}]{Lesk1986Rrot}
Lesk, A.~M., \enquote{On the calculation of Euler angles from a rotation matrix,} \emph{International Journal of Mathematical Education in Science and Technology}, Vol.~17, No.~3, 1986, pp. 335--337.
\newblock \doi{10.1080/0020739860170309}.

\end{thebibliography}

\end{document}